\documentclass[amssymb,prb,twocolumn,floats]{revtex4b3}
\usepackage{bm}
\usepackage{graphicx}
\begin{document}
\preprint{July 11, 2000}
\title{Hole-mediated ferromagnetism in tetrahedrally coordinated
semiconductors}
\author{T. Dietl}
\email{dietl@ifpan.edu.pl}
\homepage{www.ifpan.edu.pl/SL-2/sl23.htlm}
\affiliation{Research
Institute of Electrical Communication, Tohoku University, Katahira
2-1-1, Sendai 980-8577, Japan and Institute of Physics and College
of Science, Polish Academy of Sciences,  al. Lotnik\'{o}w 32/46,
PL-00-668 Warszawa, Poland}
 \author{H. Ohno and F. Matsukura}
 \affiliation{Laboratory for
Electronic Intelligent Systems, Research Institute of Electrical
Communication, Tohoku University, Katahira 2-1-1, Sendai 980-8577,
Japan}
\date{\today}
\begin{abstract}
A mean field model of ferromagnetism mediated by delocalized or
weakly localized holes in zinc-blende and wurzite diluted magnetic
semiconductors is presented. The model takes into account: (i)
strong spin-orbit and $kp$ couplings in the valence band; (ii) the
effect of strain upon the hole density-of-states, and (iii) the
influence of disorder and carrier-carrier interactions,
particularly near the metal-to-insulator transition. A quantitative
comparison between experimental and theoretical results for
(Ga,Mn)As demonstrates that theory describes the values of the
Curie temperatures observed in the studied systems as well as
explain the directions of the easy axis and the magnitudes of the
corresponding anisotropy fields as a function of biaxial strain.
Furthermore, the model reproduces unusual sign, magnitude, and
temperature dependence of magnetic circular dichroism in the
spectral region of the fundamental absorption edge. Chemical trends
and various suggestions concerning design of novel ferromagnetic
semiconductor systems are described.
\end{abstract}

\maketitle

\section{INTRODUCTION}

The discovery of ferromagnetism in zinc-blende\cite{Ohno92,Ohno96a}
III-V and\cite{Haur97,Ferr99} II-VI Mn-based compounds allows one
to explore physics of previously not available combinations of
quantum structures and magnetism in semiconductors. For instance, a
possibility of changing the magnetic phase by light in\cite{Kosh97}
(In,Mn)As/(Al,Ga)Sb and\cite{Haur97} (Cd,Mn)Te/(Cd,Zn,Mg)Te
heterostructures was put into the evidence. The injection of
spin-polarized carriers from (Ga,Mn)As to a (In,Ga)As quantum well
in the absence of an external magnetic field was demonstrated,
too.\cite{Ohno99b} It is then important to understand the
ferromagnetism in these semiconductors, and to ask whether the
Curie temperatures $T_C$ can be raised to above 300 K from the
present 110 K observed for
Ga$_{0.947}$Mn$_{0.053}$As.\cite{Ohno96a,Mats98}

In this paper, we develop theory of the hole-mediated
ferromagnetism in tetrahedrally coordinated semiconductors along
the lines of a model proposed recently by us.\cite{Diet00} Since we
aim at quantitative description of experimental findings, the
proposed theoretical approach\cite{Diet00} makes use of empirical
facts and parameters wherever possible. We begin the present paper
by discussing electronic states in p-type magnetic semiconductors.
We classify the studied systems as charge transfer insulators, so
that our theory is not applicable to materials in which d electrons
participate in charge transport. We note that Mn ions act as both
source of localized spins and effective mass acceptors. We adapt,
therefore, the physics of the metal-insulator transition in doped
semiconductors for the studied case, and assume that over the
relevant range of impurity concentrations, the ferromagnetic
exchange is mediated by delocalized or weakly localized holes.
Since the resulting spin-spin coupling is long range, we use a mean
field approximation to determine various thermodynamic,
magnetoelastic, and optical properties of the system. Particular
attention is paid to take carefully into account the complex
structure of the valence band. We then present results of
comprehensive numerical studies which provide qualitative, and in
many cases quantitative, interpretation of experimental finding
accumulated over the recent years for (Ga,Mn)As. This good
agreement between experimental and theoretical data encourages us
to show, in the final part of the paper, expected chemical trends,
and to propose various suggestions concerning the design of novel
ferromagnetic semiconductor systems.

In general terms, our results point to the importance of the $kp$
and spin-orbit interactions in the physics of the hole-mediated
ferromagnetism is semiconductors. These interactions control the
magnitude of Curie temperature, saturation value of magnetization,
and  character of magnetic anisotropies. A comparison of
theoretical and experimental findings not only emphasizes
similarities and differences  between III-V magnetic semiconductors
and other ferromagnetic systems, but demonstrates also novel
aspects of half metallic ferromagnets. The recent comprehensive
reviews present many aspects of III-V,\cite{Ohno98}
II-VI,\cite{Cibe99} as well as of IV-VI magnetic
semiconductors,\cite{Stor97} which will not be discussed here.

\section{ELECTRONIC STATES IN p-TYPE MAGNETIC SEMICONDUCTORS}
\subsection{Mn ion in tetrahedrally coordinated semiconductors}

We consider zinc-blende or wurzite semiconductor compounds, in
which the cations are partly substituted by magnetic ions, such as
Mn. The magnetic ions are assumed to be randomly distributed over
the cation sites, as found by extended x-ray absorption fine
structure (EXAFS) studies in the case of
Ga$_{1-x}$Mn$_x$As\cite{Shio98} and
Cd$_{1-x}$Mn$_x$Te.\cite{Balz85} The Mn provides the localized spin
$S = 5/2$ and, in the case of III-V semiconductors, acts as an
acceptor. These Mn acceptors compensate the deep antisite donors
commonly present in GaAs grown by low-temperature molecular beam
epitaxy, and produce a p-type conduction with metallic resistance
for the Mn concentration $x$ in the range $0.04 \le x \le
0.06$.\cite{Mats98,Oiwa97,Esch97,Shim99} Accurate analysis of the
transport data, complicated by a large magnitude of the
extraordinary Hall effect, confirms the presence of a strong
compensation,\cite{Omiy00} presumably by As antisite donors, as
mention above. However, at this stage we cannot exclude the
existence of self compensation mechanisms, such as the formation of
Mn AX-like centers or donor defects once the Fermi level reaches an
appropriately deep position in the valence band.\cite{Walu88}

According to optical studies, Mn in GaAs forms an acceptor center
characterized by a moderate binding energy\cite{Linn97} $E_a = 110$
meV, and a small magnitude of the energy difference between the
triplet and singlet state of the bound hole\cite{Linn97,Aver87}
$\Delta \epsilon = 8 \pm 3$ meV. This small value demonstrates that
the hole introduced by the divalent Mn in GaAs does not reside on
the d shell or forms a Zhang-Rice-like singlet,\cite{Zhan88,Beno92}
but occupies an effective mass Bohr orbit.\cite{Diet00,Bhat00}
Thus, due to a large intra-site correlation energy $U$, (Ga,Mn)As
can be classified as a charge-transfer insulator, a conclusion
consistent with photoemission spectroscopy.\cite{Okab98,Okab99} At
the same time, the p-d hybridization results in a spin-dependent
coupling between the holes and the Mn ions, $H_{pd} = - \beta
N_o{\bm{sS}}$. Here $\beta$ is the p-d exchange integral and $N_o$
is the concentration of the cation sites. The analysis of both
photoemission data\cite{Okab98,Okab99} and magnitude\cite{Bhat00}
of $\Delta \epsilon$ leads to the exchange energy $\beta N_o
\approx -1$~eV. Similar values of $\beta N_o$ are observed in II-VI
diluted magnetic semiconductors with comparable lattice
constants.\cite{Diet94} This confirms Harrison's suggestion that
the hybridization matrix elements depend primarily on the
inter-atomic distance.\cite{Harr87} According to the model in
question, the magnetic electrons remain localized at the magnetic
ion, so that they do not contribute to charge transport. This
precludes Zener's double exchange\cite{Akai98} as the mechanism
leading to ferromagnetic correlation between the distant Mn spins.
At the same time, for some combinations of transition metals and
hosts, the "chemical" and exchange attractive potential introduced
by the magnetic ion can be strong enough to bind the hole on a
local orbit.\cite{Zhan88,Beno92} In an intermediate regime, the
probability of finding the hole around the magnetic ion is
enhanced, which results in the apparent increase of $|\beta N_o|$
with decreasing $x$.\cite{Beno92}

\subsection{Two-fluid model of electronic states near
the metal-insulator transition}
\label{sec:two}

Ionized impurity and magnetic scatterings lead to localization of
the effective mass holes introduced by Mn in III-V compounds or by
acceptors in the case of II-VI materials. It is, therefore,
important to discuss the effect of Anderson-Mott localization on
the onset of ferromagnetism. The two-fluid model\cite{Paal91}
constitutes the established description of electronic states in the
vicinity of the Anderson-Mott metal-insulator transition (MIT) in
doped semiconductors. According to that model, the conversion of
itinerant electrons into singly occupied impurity states with
increasing disorder occurs gradually, and begins already on the
metal side of the MIT. This leads to a disorder-driven static phase
separation into two types of regions: one populated by electrons in
extended states, and another that is totally depleted from the
electrons or contains singly occupied impurity-like states. The
latter controls the magnetic response of doped non-magnetic
semiconductors\cite{Paal91} and gives rise to the presence of bound
magnetic polarons (BMP) on both sides of the MIT in magnetic
semiconductors.\cite{Diet94,Sawi86,Glod94} Actually, the formation
of BMP shifts the MIT towards the higher carrier
concentrations.\cite{Diet94,Sawi86,Glod94} On crossing the MIT, the
extended states become localized. However, according to the scaling
theory of the MIT, their localization radius $\xi$ decreases rather
gradually from infinity at the MIT towards the Bohr radius deep in
the insulator phase, so that on a length scale smaller than $\xi$
the wave function retains an extended character. Such weakly
localized states are thought to determine the static longitudinal
and Hall conductivities of doped semiconductors. The central
suggestion of the recent model\cite{Diet00} is that the holes in
the extended or weakly localized states mediate the long-range
interactions between the localized spins on both sides of the MIT
in the III-V and II-VI magnetic semiconductors.

As will be discussed below, the Curie temperature $T_C$ is
proportional to the thermodynamic spin density-of-states $\rho_s$
which, in turn, is proportional to the spin susceptibility of the
carrier liquid $\chi_s$. Like other thermodynamic quantities,
$\rho_s$ does not exhibit any critical behavior at the MIT.
However, $\rho_s$ exhibits large space fluctuations at criticality,
which will result in local fluctuations of magnetic properties. The
quantitative renormalization of $\rho_s$ by disorder will depend on
its microscopic nature, for instance, on the degree of
compensation. The enhancement of $\rho_s$ by the carrier-carrier
interactions can be described by the Fermi-liquid parameter $A_F$,
$\rho_s \rightarrow A_F\rho_s$.\cite{Diet97} The value of $A_F =
1.2$, as evaluated\cite{Jung99} by the local-spin-density
approximation for the relevant hole concentrations, has been
adopted for our computations. We note that disorder-modified
carrier-carrier interactions in the triplet particle-hole channel
tend to enhance $A_F$, which may even drive the system towards a
Stoner-like instability.\cite{Beli94} In magnetic systems, however,
spin-flip scattering by fluctuations of magnetization makes this
enhancement mechanism rather inefficient.

The two-fluid model is consistent with the recent EPR
results,\cite{Szcz99b} which point the coexistence of the neutral
Mn acceptors and ionized Mn d$^5$ states in some range of Mn
concentration. Furthermore, the observed\cite{Szcz99a,Besc99} sign
of magnetic circular dichroism in (Ga,Mn)As suggests the presence
of the Fermi-liquid like states on the both sides of the MIT, as we
shall discuss in Sec.~\ref{sec:mcd}.

\subsection{Valence band structure and exchange splitting
of the hole subbands}

Since the valence band originates merely from the anion p and
cation d wave functions, the exchange interaction mediated by the
holes is expected to be strongly affected by anisotropy of hole
dynamics and the coupling between the spin and orbital orbit
degrees of freedom. To take those effects into account, the hole
dispersion and wave functions are computed by diagonalizing the 6x6
Kohn-Luttinger $kp$ matrix.\cite{Bir74} In this model, four
$\Gamma_8$ and two $\Gamma_7$ bands are taken explicitly into
account, whereas other bands are included by second order
perturbation theory.  The model is developed for zinc-blende and
wurzite semiconductors. It allows for warping, quantizing magnetic
fields, and biaxial strain but no terms associated with the lack of
the symmetry inversion are taken into account. The effect of the
spin-dependent interaction between the holes and the Mn spins is
described in terms of the virtual-crystal and molecular-field
approximations,\cite{Gaj78} so that
\begin{equation}
H_{pd} = \beta {\bm {sM}}({\bm{r}})/g\mu_B,
 \label{eq:pd}
\end{equation}
where ${\bm M}({\bm r})$ is the magnetization of the localized
spins that carry magnetic moment $-Sg\mu_B$, where $S = 5/2$ and $g
=2.0$.

The explicit form of the $kp$ 6x6 matrices, together with the
matrix $H_{pd}$ derived by us in the Luttinger-Kohn representation
for arbitrary directions of magnetization $\bm{ M}$, are displayed
in Appendix A. We note that for the case under consideration,
involving large exchange interaction and high kinetic energy,
particularly important are off diagonal terms describing the p-d
coupling between the $\Gamma_8$ and $\Gamma_7$ bands. A numerical
procedure that serves to determine the concentration, free energy,
wave functions, and optical characteristics of the holes is
outlined in Appendix B. The adopted values of the Luttinger
parameters $\gamma_i$ and the spin-orbit splittings $\Delta_o$ are
summarized in Appendix C for various III-V and II-VI parent
compounds. According to photoemission studies\cite{Okab98} $\beta
N_o = -1.2 \pm 0.2$ eV for (Ga,Mn)As. The value $\beta N_o = -1.2$
eV is taken for the computations, though future works may reveal
some dependence of $\beta$ of Eq.~(\ref{eq:pd}) on Mn and/or hole
concentrations because of energetic proximity of the Fermi and
relevant Mn d-levels. Such a dependence could result also from
corrections to the virtual-crystal and molecular-field
approximations if a short-range part of the Mn potential is
attractive for the valence band hole.\cite{Beno92}

\begin{figure}
\includegraphics*[width=90mm]{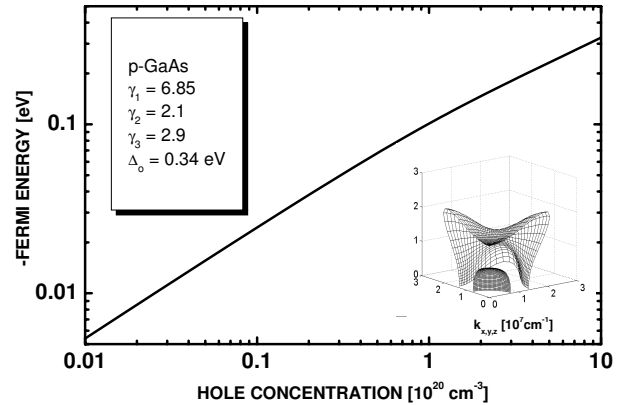}
 \caption[]{Fermi energy as a function of the hole concentration $p$
computed from the 6x6 Luttinger model for $kp$ parameters displayed
in the figure. Inset shows cross section of the Fermi sphere for
$p=3.5\times 10^{20}$ cm$^{-3}$.} \label{fig:Evsp_1}
\end{figure}

Figure \ref{fig:Evsp_1} presents the dependence of the low
temperature Fermi energy $\varepsilon_F$ on the hole concentration
$p$ computed for the band parameters corresponding to GaAs. No
effects of p-d exchange, disorder, or Coulomb interactions between
the holes are taken into account. The corresponding
density-of-states (DOS) effective mass increases from 0.67$m_o$ in
the limit of small hole concentrations to the value of 1.35$m_o$
for $p = 5\times 10^{20}$ cm$^{-3}$. This increase is caused by the
$kp$ interaction between the $\Gamma_8$ and $\Gamma_7$ bands,
which--because of a relatively small magnitude of $\Delta_o$ in
GaAs--is important for the relevant hole densities. The inset to
Fig.~\ref{fig:Evsp_1} shows the cross section of the Fermi surface
for $p=3.5\times 10^{20}$ cm$^{-3}$, which corresponds to
$\varepsilon_F = -195$ meV in respect to energy of the $\Gamma_8$
point. Two valleys, the heavy and light hole subbands, are visible.

A strong and complex influence of the p-d interaction and strain
upon the valence band is shown in Fig.~\ref{fig:fs_2}. The cross
sections of the Fermi spheres are depicted for $p=3.5\times
10^{20}$ cm$^{-3}$, which now corresponds to $\varepsilon_F \approx
-165$ meV,  and for the parameter of the exchange splitting,
\begin{equation}
B_G = A_F\beta M/6g\mu_B, \label{eq:bg}
\end{equation}
taken as $B_G = -30$ meV. If $\beta N_o = -1.2$ eV and $A_F = 1.2$,
this magnitude of $B_G$ occurs for the saturation value of
magnetization $M$ at $x= 0.05$. We note at this point that some of
the effects, e.g., the direction of spin-polarization vector
depends on the sign of $\beta$ and thus $B_G$, whereas others, like
the Curie temperature, are proportional to $\beta^2$.  In the
absence of strain, $\epsilon =0$, the four fold degeneracy at the
$\Gamma$ point is lifted by the p-d exchange, and the corresponding
energies at $k=0$ are $\pm 3B_G$ and $\pm B_G$ for $\Delta_o \gg
|B_G|$. However, the splitting at non-zero wave vectors depends on
the relative orientation of ${\bm M}$ and ${\bm k}$.\cite{Gaj78} In
particular, since the spin of the heavy hole is polarized along
${\bm k}$ for $\epsilon =0$, the exchange splitting is seen to
vanish for ${\bm k} \perp {\bm M}$. This mixing of orbital and spin
degrees of freedom, together with highly nonparabolic, anisotropic,
and mutually crossing dispersion relations constitute important
aspects of the hole-mediated ferromagnetism in terahedrally
coordinated semiconductors.

\section{HOLE-INDUCED FERROMAGNETISM IN SEMICONDUCTORS}

\subsection{Short-range atiferromagnetic superexchange and
ferromagnetic double exchange}

In addition to the interaction between the carriers and localized
spins, the p-d hybridization leads also to the superexchange, a
short-range antiferromagnetic coupling between the Mn spins. The
superexchange is mediated by spin-polarization of occupied electron
bands, in contrast to the Zener ferromagnetic exchange which is
mediated by spin-polarization of the carrier liquid. The
antiferromagnetic exchange dominates in undoped II-VI
semiconductors,\cite{Diet94} and also in a compensated
Ga$_{1-x}$Mn$_x$As:Sn.\cite{Nish97} In order to take the influence
of this interaction into account, it is convenient to parameterize
the dependence of magnetization on the magnetic field in the
absence of the carriers, $M_o(H)$, by the Brillouin function
${\mbox{B}}_S$ according to
\begin{equation}
M_o(H) = g\mu_BSN_ox_{eff}{\mbox{B}}_S[g\mu_BH/k_B(T+T_{AF})],
\label{eq:br}
\end{equation}
where two empirical parameters, the effective spin concentration
$x_{eff}N_o < xN_o$ and temperature $T_{eff} > T$, take the
presence of the superexchange interactions into
account.\cite{Diet94,Shap86} The dependencies $x_{eff}(x)$ and
$T_{AF}(x)\equiv T_{eff}(x) - T$ are known\cite{Twar84a,Shap86} for
II-VI compounds.

In order to elucidate the effect of doping on $x_{eff}$ and
$T_{AF}$ we refer to the two-fluid model described in
Sec.~\ref{sec:two}. In terms of that model the delocalized or
weakly localized holes account for the ferromagnetism. Actually,
the participation of the same set of holes in both charge transport
and the ferromagnetic interactions is shown, in
(Ga,Mn)As\cite{Oiwa97} and in (Zn,Mn)Te,\cite{Ferr00} by the
agreement between the temperature and field dependencies of the
magnetization deduced from the extraordinary Hall effect, $M_H$,
and from direct magnetization measurements, $M_D$, particularly in
the vicinity of $T_C$. However, below $T_C$ and in the magnetic
fields greater than the coercive force, while $M_H$ saturates (as
in standard ferromagnets), $M_D$ continues to rise with the
magnetic field.\cite{Oiwa97,Besc99} Since $M_H$ is proportional to
spin polarization of the carriers, its saturation may reflect a
saturation of hole polarization, which--for appropriately low
values of the Fermi energy--can occur even if the Mn spins are not
totally spin polarized (half metallic case). It appears, however,
that Mn spins in the regions depleted from the carriers (which,
therefore, do not participate in the long range magnetic order)
contribute also to a slowly increasing component of $M_D(H)$.

According to the two fluid model, part of the carriers are trapped
on strongly localized impurity states, and thus form bound magnetic
polarons (BMP). If there is an exchange coupling between the two
fluids, the BMP participate in the formation of the ferromagnetic
order. Furthermore, the coupling between the BMP appears to be
ferromagnetic, at least in some range of relevant
parameters.\cite{Wolf96,Bhat99} To gain the Coulomb energy, the BMP
are preferentially formed around close pairs of ionized acceptors.
In the case of III-V materials this leads, {\it via} Zener's double
exchange,\cite{Zene50} to a local ferromagnetic alignment of
neighbor Mn d$^5$ negative ions,\cite{Blin97} so that $x \approx
x_{eff}$ and $T_{AF} \approx 0$. By contrast, in II-VI compounds,
for which acceptor cores do not carry any spin and the degree of
compensation is low, BMP are not preferentially formed around Mn
pairs, so that the close pairs remained antiferromagnetically
aligned, even in p-type samples. The presence of a competition
between the ferromagnetic and antiferromagnetic interactions in
II-VI compounds, and its absence in III-V materials, constitutes
the important difference between those two families of magnetic
semiconductors.

\subsection{Zener model of ferromagnetic interactions mediated
by free carriers}

As mentioned above, we assume that weakly localized or delocalized
holes mediate long-range ferromagnetic interaction between the
spins. Zener\cite{Zene50} first proposed a model of ferromagnetism
driven by the exchange coupling of the carriers and the localized
spins. According to that model, spin-polarization of the localized
spins leads to spin-splitting of the bands, which results in the
lowering of the carrier energy. At sufficiently low temperature,
this lowering overcompensates the increase of the free energy
caused by a decrease of entropy, which is associated with the
polarization of the localized spins.  However, the Zener model was
later abandoned, as neither the itinerant character of the magnetic
electrons nor the quantum (Friedel) oscillations of the electron
spin-polarization around the localized spins were taken into
account, both of these are now established to be critical
ingredients for the theory of magnetic metals. In particular, a
resulting competition between ferromagnetic and antiferromagnetic
interactions in metals leads rather to a spin-glass than to a
ferromagnetic ground state. In the case of semiconductors, however,
the mean distance between the carriers is usually much greater than
that between the spins. Under such conditions, the exchange
interaction mediated by the carriers is ferromagnetic for most of
the spin pairs, which reduces the tendency towards spin-glass
freezing. Actually, for a random distribution of the localized
spins, the mean-field value of the Curie temperature $T_C$ deduced
from the Zener model is equal to that obtained from the Ruderman,
Kittel, Kasuya, and Yosida (RKKY)
approach,\cite{Diet97,Diet99,Kitt68,Lars81} in which the presence
of the Friedel oscillations is explicitly taken into account.

\subsection{Mean-field model of Curie temperature and thermodynamic
properties}

As reported elsewhere,\cite{Diet00} the Zener model describes
correctly the experimental values of $T_C$ in both (Ga,Mn)As and
(Zn,Mn)Te, provided that band structure effects are taken into
account. The starting point of the model is the determination how
the Ginzburg-Landau free-energy functional $F$ depends on the
magnetization $M$ of the localized spins. The hole contribution to
$F$, $F_c[M]$ is computed by diagonalizing the 6x6 Kohn-Luttinger
matrix together with the p-d exchange contribution, and by the
subsequent computation of the partition function $Z$, as described
in Appendix B. This model takes the effects of the spin-orbit
interaction into account, a task difficult technically within the
RKKY approach, as the spin-orbit coupling leads to non-scalar terms
in the spin-spin Hamiltonian. Moreover, the indirect exchange
associated with the virtual spin excitations between the valence
subbands, the Bloembergen–-Rowland mechanisms,\cite{Diet94} is
automatically included. The remaining part of the free energy
functional, that of the localized spins, is given by
\begin{equation}
F_S[M] = \int_0^M d M_o H(M_o), \label{eq:fs}
\end{equation} where
$H(M_o)$ is given in Eq.~(\ref{eq:br}). By minimizing $F[M] =
F_c[M] + F_S[M]$ with respect to $M$ at given $T$, $H$, and hole
concentration $p$, one obtains $M(T,H)$ as a solution of the
mean-field equation,
\begin{equation}
M = \mu_g\mu_BSN_ox_{eff}{\mbox{B}}_S\left[\frac{g\mu_B(-\partial
F_c[M]/\partial M + H)}{k_B(T+T_{AF})}\right].
\end{equation}

Near the Curie temperature $T_C$, where $M$ is small, we expect
$F_c[M] - F_c[0] \sim M^2$. It is convenient to parameterize this
dependence by a spin density-of-states  $\rho_s$,
\begin{equation}
F_c[M] = F_c[0] - A_F\rho_s\beta^2M^2/2(2g\mu_B)^2.
 \label{eq:fc}
\end{equation}
The spin density-of-states $\rho_s$ is related to carrier magnetic
susceptibility according to $\chi = A_F(g^*\mu_B)^2\rho_s/4$  and,
in general, has to be determined numerically by computing $F_c[M]$.
By expanding B$_S(M)$ we arrive to,
\begin{equation}
T_C = x_{eff}N_oS(S+1)\beta^2A_F\rho_s(T_C)/12k_B - T_{AF}.
\label{eq:tc}
\end{equation}
For a strongly degenerate carrier liquid |$\varepsilon_F|/k_BT \gg
1$, as well as neglecting the spin-orbit interaction, $\rho_s$
becomes equal to the total density-of-states $\rho$ for intra-band
charge excitations, where $\rho = m^*_{DOS}k_F/\pi^2\hbar^2$.

\begin{figure}
\includegraphics*[width=90mm]{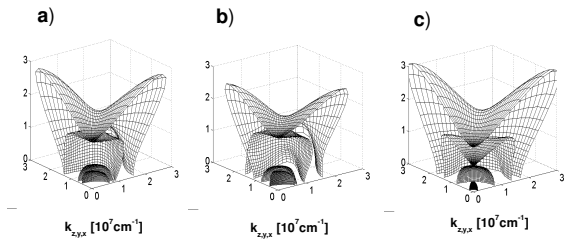}
\caption[]{Cross sections of the Fermi spheres for saturation value
of magnetization $M$ in Ga$_{095}$Mn$_{0.05}$As with hole
concentration $p=3.5\times 10^{20}$ cm$^{-3}$.  Results for various
orientations of magnetization ${\bm M}$ and biaxial strains
$\epsilon_{xx}$ are shown: (a) $M\parallel$ [100]; $\epsilon_{xx}
=0$. (b) $M\parallel$ [100]; $\epsilon_{xx} = -2$\%.
(c)$M\parallel$ [001]; $\epsilon_{xx} =+2$\%. A strong dependence
of the splitting on the relative orientation of the wave vector and
magnetization is visible.} \label{fig:fs_2}
\end{figure}

\begin{figure}
\includegraphics*[width=90mm]{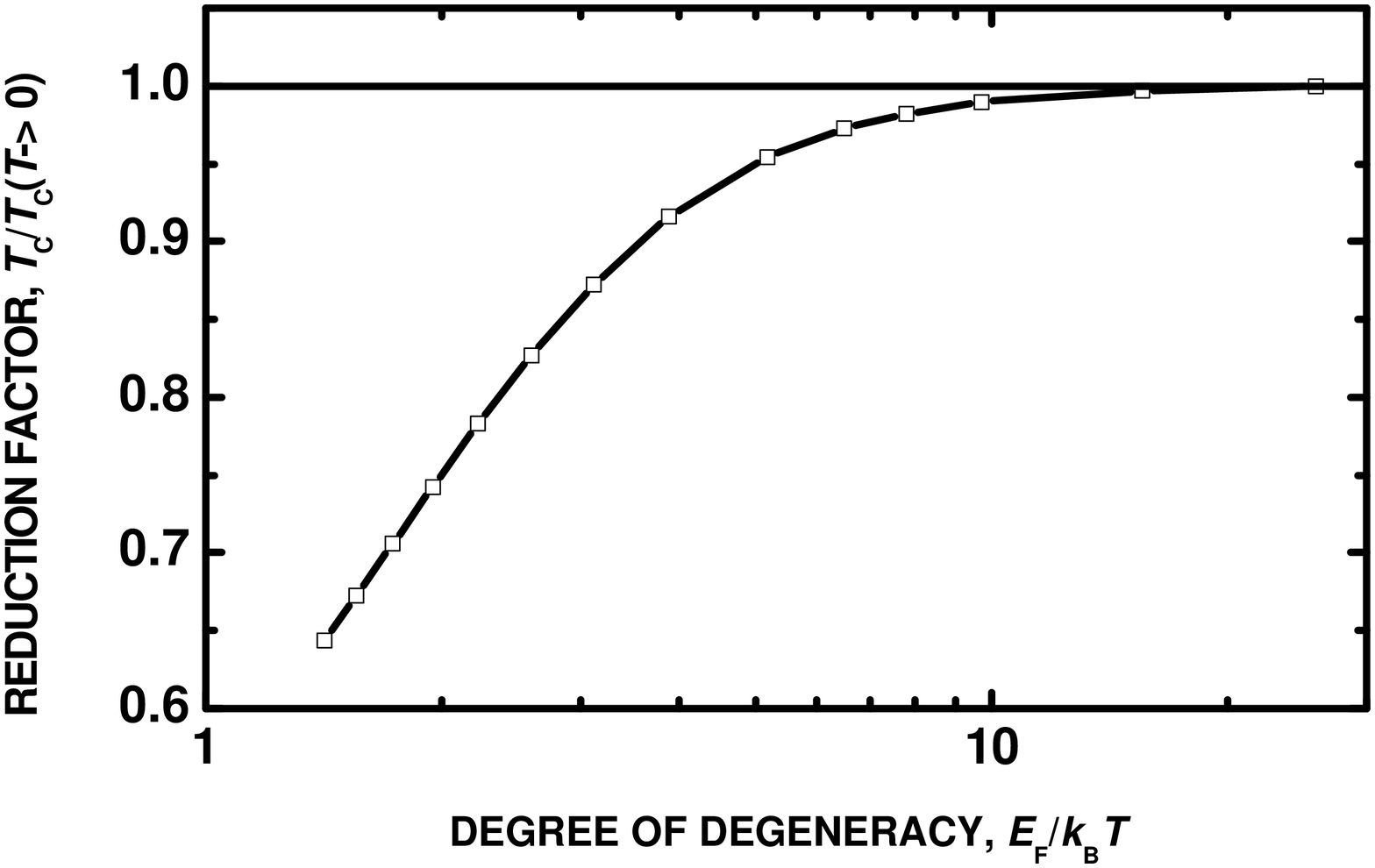}
\caption[]{The computed ratio the actual Curie temperature $T_C$ to
that obtained assuming that the hole liquid is strongly degenerate
as a function of the degeneration factor.}
 \label{fig:dege_3}
\end{figure}

 In order to check the quantitative significance of
carrier entropy, the computed values of $T_C$ were compared to
those obtained assuming strong degeneration of the carrier liquid.
As shown in Fig.~\ref{fig:dege_3} such an assumption leads to error
smaller than 1\% if $\varepsilon_F|/k_BT > 10$. Thus, in this
range, the carrier energy, not free energy, was used for the
evaluation of $T_C$. Furthermore, we take $x_{eff} = x$ and $T_{AF}
= 0$, as discussed above.

\begin{figure}
\includegraphics*[width=90mm]{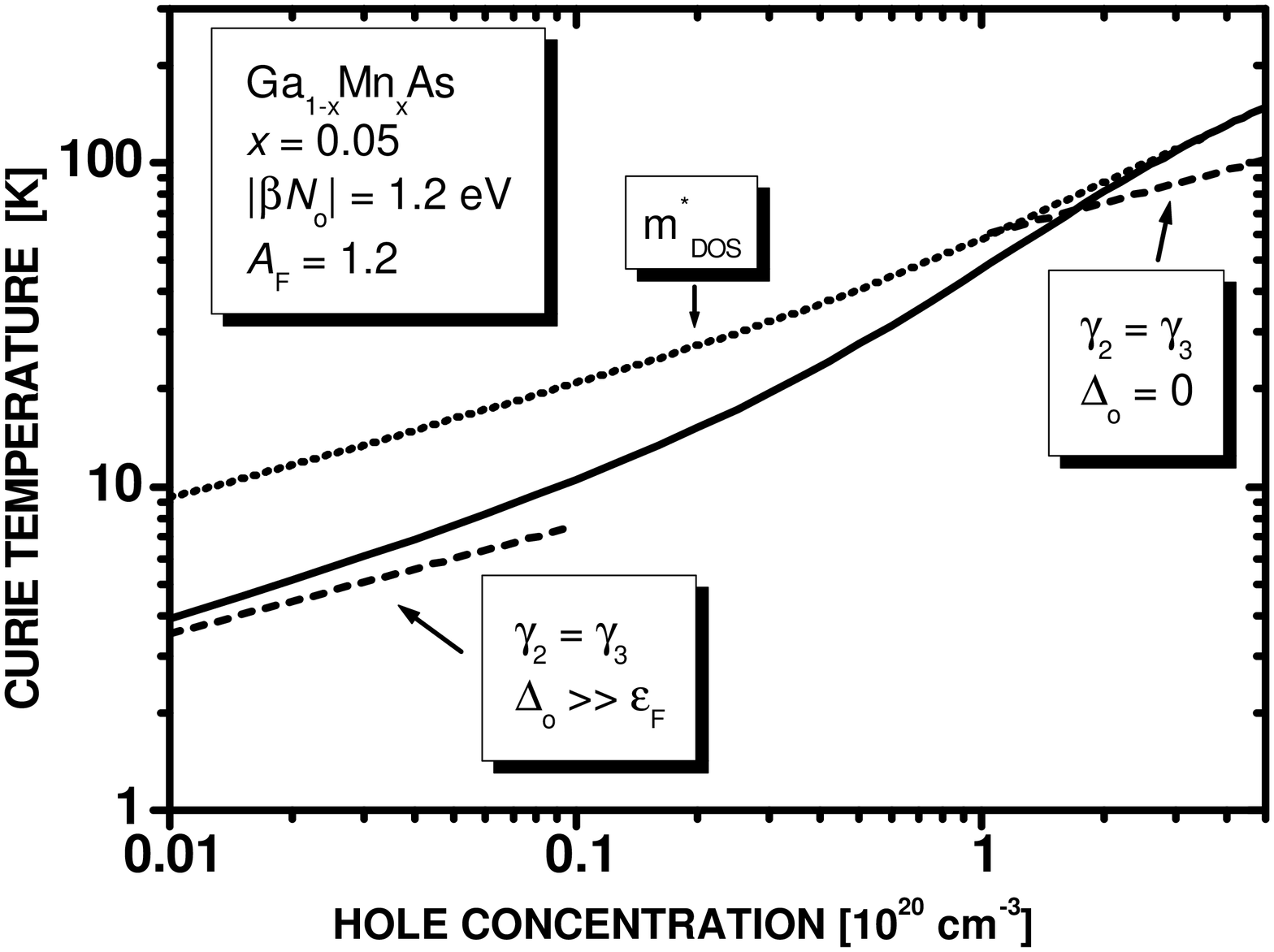}
\caption[]{Curie temperature as a function of the hole
concentration for Ga$_{0.95}$Mn$_{0.05}$As computed from the 6x6
Luuttinger model (solid line). Straight dashed lines represent
results obtained assuming a large and small value of the spin-orbit
splitting $\Delta_o$, respectively. The dotted line is calculated
neglecting the effect of the spin-orbit interaction on the hole
spin susceptibily.}
 \label{fig:tc_4}
\end{figure}

The values of $T_C$ as a function of the hole concentration $p$ for
Ga$_{0.95}$Mn$_{0.05}$As, as computed by our model, are shown by
the solid line in Fig.~\ref{fig:tc_4}. Since the magnitude of $T_C$
is directly proportional to $x_{eff}N_o\beta^2A_F$, the theoretical
results can be easily extended to other values of $x$, $\beta$ or
$A_F$. In particular, $T_C=300$ K is expected for $x=0.125$ and
$p=3.5\times 10^{20}$ cm$^{-3}$.

\subsection{Validity of the model and comparison to other approaches}

Equation (\ref{eq:tc}) with $\rho_s = \rho$, that is neglecting
effects of the spin-orbit interaction, has already been derived by
a number of equivalent methods.\cite{Diet94,Diet97,Diet99,Jung99}
In the present work $F_c[M]$, which served to determine $T_C$,
$M(T,H)$, anisotropy fields, and optical spectra was obtained from
the 6x6 $kp$ model, as described in Appendices A and B. In order to
illustrate the importance of band structure effects, the evaluation
of $T_C$ has been performed employing various models.  The two
straight dashed lines in Fig.~\ref{fig:tc_4} depict the expected
$T_C(p)$ if the warping is neglected ($\gamma_2 = \gamma_3$) as
well as a large and small spin-orbit splitting
$E_{\Gamma_8}-E_{\Gamma_7} \equiv \Delta_o$ is assumed,
respectively. We see that the above assumptions are approximately
fulfilled in low and high concentration range, respectively, but
the calculation within the full model is necessary in the
experimentally relevant region of the intermediate hole densities.
In particular, 4x4 model\cite{Ferr00,Abol00} ($\Delta_o \gg
|\varepsilon_F|$) would be reasonably correct for (Zn,Mn)Te, where
$\Delta_o$ is about 1 eV but not for (Ga,Mn)As, for which
$\Delta_o$ is almost three times smaller. Finally, the dotted line
shows $T_C$ evaluated replacing $\rho_s$ by $\rho$. We see that
$\rho \approx \rho_s$ already for $p \approx 10^{20}$ cm$^{-3}$,
despite that $|\varepsilon_F|$ is still two times smaller than
$\Delta_o$. In contrast, $\rho_s$ is more than two times smaller
than $\rho$ in the limit of small $p$. This reduction of the spin
susceptibility, and thus of $T_C,$ stems from the absence of the
exchange splitting of the heavy hole subband for ${\bm{k}} \perp
{\bm{M}}$.

As described above, $T_C$ can be computed by minimizing the free
energy, and without referring to the explicit form of the
Kohn-Luttinger amplitudes $u_{i{\bm k}}$. Since near $T_C$ the
relevant magnetization $M$ is small, the carrier free energy, and
thus $T_C$, can also be determined from the linear response theory.
The corresponding $\rho_s$ assumes the standard form
\begin{equation}
\rho_s = \lim_{q \rightarrow 0} 8\sum_{ij{\bm k}} \frac{|\langle
u_{i,{\bm k}}|s_M|u_{j,{\bm k}+{\bm q}}\rangle|^2 f_i({\bm k})[1 -
f_j({\bm k}+{\bm q})]} {E_j({\bm k}+{\bm q})- E_i({\bm k})},
\label{eq:linh}
\end{equation}
where $s_M$ is the component of spin operator along the direction
of magnetization and $f_i({\bm k})$ is the Fermi-Dirac distribution
function for the $i$-th valence band subband. The equivalence of
$\rho_s$ as given by Eqs.~(\ref{eq:fc}) and (\ref{eq:linh}) can be
checked for the 4x4 spherical model by using the explicit
form\cite{Szym78} of $E_i(k)$ and $u_i$. Such a comparison
demonstrates that almost a 30\% contribution to $T_C$ originates
from interband polarization.

It is straightforward to generalize the model for the case of the
carriers confined to the $d$-dimensional
space.\cite{Diet97,Diet99,Lars81} The tendency towards the
formation of spin-density waves in low-dimensional
systems\cite{Diet99,Koss00} as well as possible spatial correlation
in the distribution of the magnetic ions can also be taken into
account. The mean-field value of the critical temperature $T_{{\bm
q}}$, at which the system undergoes the transition to a spatially
modulated state characterized by the wave vector ${\bm q}$, is
given by the solution of the equation,
\begin{equation}
\beta^2A_F({\bm q},T_{{\bm q}})\rho_s({\bm q},T_{{\bm q}}) \int d
{\bm{\zeta}}\chi_o({\bm q},T_{{\bm q}},{\bm{\zeta}})
|\phi_o({\bm{\zeta}})|^4 = 4g^2\mu_B^2.
\end{equation}
Here ${\bm q}$ spans the $d$-dimensional space,
$\phi_o({\bm{\zeta}})$ is the envelope function of the carriers
confined by a $(3 - d)$-dimensional potential well
$V({\bm{\zeta}})$; $g$ and $\chi_o$ denote the Land\'e factor and
the ${\bm q}$-dependent magnetic susceptibility of the magnetic
ions in the absence of the carriers, respectively. Within the
mean-field approximation (MFA), such magnetization shape and
direction will occur in the ordered phase, for which the
corresponding $T_{{\bm q}}$ attains the highest value. A
ferromagnetic order is expected in the three dimensional (3D) case,
for which a maximum of $\rho_s({\bm q})$ occurs at $q = 0$.

Since within the MFA the presence of magnetization fluctuations is
neglected, our model may lead to an overestimation of the magnitude
of $M$, and thus of $T_C$. To address this question we recall that
the decay of the strength of the carrier-mediated exchange
interaction with the distance between the spins $r$ is described by
the RKKY function, which in the 3D situation assumes the
form,\cite{Diet99,Kitt68,Lars81}
\begin{equation}
J(r) \sim [\sin(2k_Fr) - 2k_Fr\cos(2k_Fr)]/(2k_Fr)^4.
\end{equation}
In the case of semiconductors an average distance between the
carriers $r_c = (4\pi p/3)^{-1/3}$ is usually much greater than that
between the spins $r_S = (4\pi xN_o/3)^{-1/3}$. This means that the
carrier-mediated interaction is ferromagnetic and effectively long
range for most of the spins as the first zero of the RKKY function
occurs at $r \approx 1.17r_c$. A theoretical study\cite{Fish72} of
critical exponents for a $d$-dimensional space showed that as long
as $\sigma < d/2$ in the dependence $J(r) \sim 1/r^{d+\sigma}$, the
mean-field approach to the long wavelength susceptibility $\chi(T)$
is valid, a conclusion not affected presumably by disorder in the
spin distribution. At the same time, both relevant length scale
$r_c$, not $r_S$, and the critical exponents\cite{Fish72} $\eta =
2-\sigma$ and $\nu = 1/\sigma$, point to much faster decay of
$\chi(q)$ with $q$ than that expected from the classical
Ornstein-Zernike theory.\cite{Yeom93} This indicates that the MFA
should remain valid down to at least $|T-T_C|/T_C \approx r_S/r_c
\ll 1$. Actually, the decay of $\chi(q)$ with $q$ in the range $0 <
q < 2k_F$ is corroborated by the observation of smaller critical
scattering of the holes by magnetization fluctuations than that
calculated for $\chi(q) = \chi(0)$.\cite{Omiy00}

Recently, Monte-Carlo studies of carrier-mediated ferromagnetism in
semiconducors have been initiated.\cite{Nish98,Boss00} Such an
approach has a potential to test the accuracy of the MFA and to
determine the actual spin configuration corresponding to the ground
state. Preliminary results appear to confirm the validity of the
MFA\cite{Nish98,Boss00} as well as to indicate a possibility of the
existence of non-collinear magnetic structures in low-dimensional
systems.\cite{Boss00} Another significant recent
development\cite{Koni00} is the examination of spin-wave
excitations, their spectrum and effect on magnetization. A strong
reduction of $T_C$ was predicted,\cite{Koni00} though it should be
noted that the spin wave approximation breaks usually down at
criticality. In contrast, such an approach offers valid results at
low temperatures, provided that effects of magnetic anisotropy are
thoroughly taken into account. Furthermore, the existence of
ferromagnetic ground state was confirmed by {\it ab initio} total
energy computations for magnetic III-V
compounds.\cite{Akai98,Shir98,Kato99} However, to what extend the
employed procedures have been capable to capture accurately
correlation effects on the open d-shells seems to be unclear by
now.

Finally, we would like to stress once more that if concentrations
of the carriers and the spins become comparable, $r_c \le r_S$,
randomness associated with the competition of ferromagnetic and
antiferromagnetic interactions can drive the system towards a
spin-glass phase.\cite{Egge95} In the case of II-VI compounds, the
antiferromagnetic component is additionally enlarged by the
superexchange interaction.\cite{Ferr00} Furthermore, scattering by
ionized impurities, and the associated non-uniform distribution of
carriers in semiconductors near the MIT, may enhance disorder
effects even further. We note also that in the extreme case, $r_c
\ll r_S$, the Kondo effect, that is dynamic screening of the
localized spins by the sea of the carriers may preclude both
ferromagnetic and spin-glass magnetic ordering.

\section{COMPARISON OF THEORETICAL AND EXPERIMENTAL RESULTS FOR
(Ga,Mn)As}
\subsection{Curie temperature and spontaneous magnetization}

The most interesting property of Ga$_{1-x}$Mn$_x$As epilayers is
the large magnitude of $T_C$, up to 110 K for the Mn concentration
$x = 5.3$\%.\cite{Ohno96a,Mats98} Because of this high $T_C$, the
spin-dependent extraordinary contribution to the Hall resistance
$R_H$ persists up to 300 K, making an accurate determination of the
hole density difficult.\cite{Mats98,Oiwa97,Esch97,Shim99} However,
the recent measurement\cite{Ohno99a} of $R_H$ up to 27 T and at 50
mK yielded an unambiguous value of $p = 3.5\times10^{20}$ cm$^{-3}$
for the metallic Ga$_{0.947}$Mn$_{0.053}$As sample, in which $T_C$
= 110 K is observed.\cite{Mats98} The above value of $p$ is about
three times smaller than $xN_o$, confirming the importance of
compensation in (Ga,Mn)As. As shown in Fig.~\ref{fig:tc_4}, the
numerical results lead to $T_C = 120$ K for $x = 0.05$, and thus,
$T_C = 128$ K for $x = 0.053$ and $p=3.5\times 10^{20}$ cm$^{-3}$.
We conclude that the proposed model describes, with no adjustable
parameters, high values of $T_C$ found in (Ga,Mn)As. Furthermore,
scaling theory of electronic states near the MIT, discussed in
Sec.~\ref{sec:two}, makes it possible to explain the presence of
the ferromagnetism on the both sides of the MIT, and a non-critical
evolution of $T_C$ across the critical point.\cite{Mats98} A
comparison between theoretical and experimental data in a wider
range of Mn and hole concentrations requires reliable information
on the hole density in particular samples, which is not presently
available in the case of (Ga,Mn)As.

In the case of (Zn,Mn)Te:N, the hole concentration can readily be
determined by the Hall effect measurements at 300
K.\cite{Ferr99,Ferr00} The absence of the extraordinary Hall effect
at 300 K stems form the much lower values of $T_C$, $0.75 \le T_C
\le 2.4$ K for 5\% $\ge x \ge 2$\% and $10^{19} \le p \le10^{20}$
cm$^{-3}$. The model in question explains
satisfactorily\cite{Ferr00} $T_C(x,p)$. Three effects conspire to
make $T_C$ much greater in p-(Ga,Mn)As than in p-(Zn,Mn)Te at given
$p$ and $x$. First, as already mentioned, the small value of the
splitting $\Delta_o$ makes the reduction of $T_C$ by spin-orbit
coupling of minor importance in the case of (Ga,Mn)As. Second,
because of a smaller lattice parameter, the product $\beta N_o$ is
greater in (Ga,Mn)As. Finally, ferromagnetic double-exchange
between closely lying pairs of Mn ions is stronger than
antiferromagnetic superexchange in compensated (Ga,Mn)As. In
contrast, antiferromagnetic superexchange remains significant in
p-(Zn,Mn)Te, as the Mn atoms are electrically inactive in II-VI
compounds.

\begin{figure}
\includegraphics*[width=90mm]{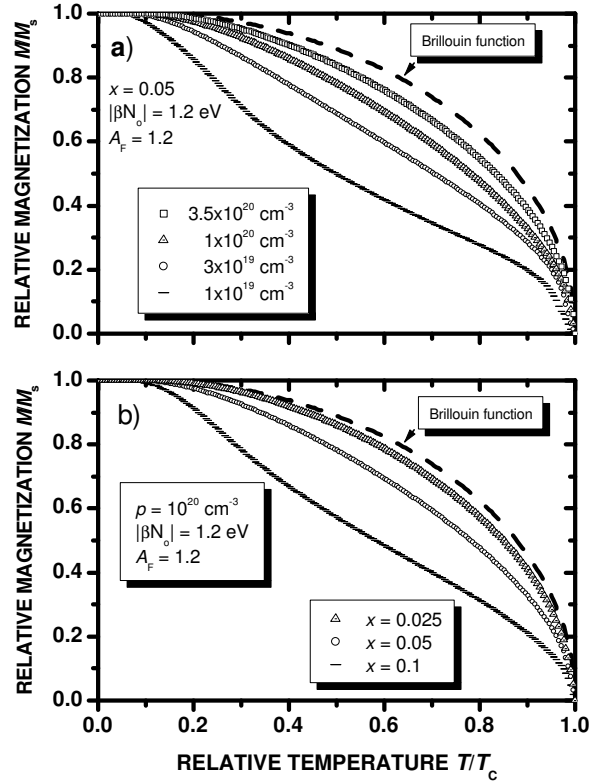}
\caption[]{The computed emerging of spontaneous magnetization $M$
below the Curie temperature $T_C$ for various hole concentrations
$p$ (a) and Mn contents $x$ (b)in Ga$_{0.95}$Mn$_{0.05}$As
(points).  If the exchange splitting of the valence band is much
smaller than the Fermi energy (large $p$, small $x$), the evolution
of magnetization follows the Brillouin function (dashed lines).}
 \label{fig:bril_5}
\end{figure}

Important characteristics of ferromagnets are the magnitude and
temperature dependence of the spontaneous magnetization $M$ below
$T_C$. Experimental studies of metallic samples indicates that on
lowering temperature $M$ increases as the Brillouin function
reaching the saturation value $M_s$ for $T \rightarrow
0$.\cite{Mats98} This behavior indicates that a molecular field
$H^*$ acting on the Mn spins is proportional to their magnetization
$M$. Figure~\ref{fig:bril_5} presents values of $M/M_s$ as a
function of $T/T_C$ calculated with no adjustable parameter for
(Ga,Mn)As containing various hole and Mn concentrations. We see
that indeed the computed Mn spin magnetization $M(T)/M_s$ follows,
to a good approximation, the Brillouin function, except for
materials with rather small values $p$ or large values of $x$. The
latter cases correspond to the half-metallic situation, for which
only the ground-state hole subband is populated even at $M < M_s$,
so that the molecular field produced by the carriers $H^*$ attains
its maximum value and, therefore, ceases to be proportional to $M$.
Nevertheless, the saturation value $M_s$ can be reached provided
that temperature is sufficiently low, $k_BT_{eff}\ll Sg\mu_BH^*$.
Such a half-metallic behavior is observed in the case of
ferromagnetic correlation imposed by a dilute hole liquid in
(Cd,Mn)Te quantum wells.\cite{Koss00}

\subsection{Hole magnetization and spin polarization}

The magnitude of magnetization presented in the previous section
was computed neglecting a possible contribution originating from
hole magnetic moments. Such a contribution can be significant as
the hole liquid is spin-polarized for non-zero magnetization of the
Mn spins. Because of the spin-orbit interaction, the hole
magnetization consists of two components. One comes from magnetic
moments of the hole spins, described by the Land\'e factor of the
free electrons and the Luttinger parameter $\kappa$. Another
contribution, absent for localized carriers, originates from
diamagnetic currents, whose magnetic moments can be oriented along
the spin polarization by the spin-orbit interaction. The evaluation
of the latter requires the inclusion of the Landau quantization in
the $kp$ hamiltonian. The carrier magnetization $M_c$ is then given
by,
\begin{equation}
M_c =-\lim_{T,H \rightarrow 0}\partial G_c(H)/\partial H,
\end{equation}
where the Gibbs thermodynamic potential is calculated for a given
value of the Mn spin magnetization $M$ and the Fermi energy
$\varepsilon_F(M)$ as a function of the magnetic field acting on
the carriers $H$.  In general, the eigenvalue problem for the holes
in the magnetic field cannot be transformed into an algebraic
equation. Such a transformation is, however, possible if the
warping is neglected. We have, therefore, calculated the hole
magnetization $M_c$ disregarding the anisotropy, that is assuming
$\gamma_1 = 6.85$, $\gamma_2 = \gamma_3= 2.58$, and $\kappa =1.2 $.
The explicit form of the relevant Luttinger matrices is displayed
in Appendix A. The partition function $Z$ was computed by summing
over the Landau index $n$, the wave vector $k_z$, and the six hole
subbands.

\begin{figure}
\includegraphics*[width=90mm]{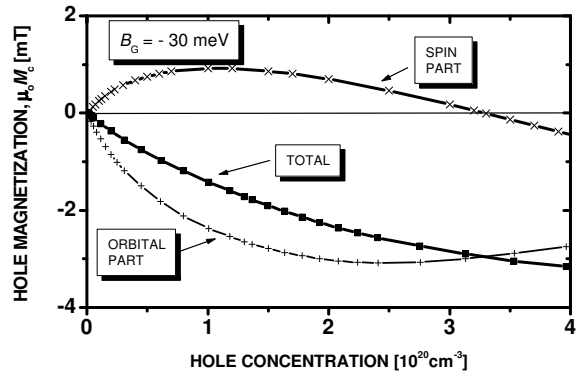}
\caption[]{Magnetization $M_c$ of the hole liquid (squares) in
Ga$_{1-x}$Mn$_{x}$As computed as a function of the hole
concentration for the spin splitting parameter $B_G = -30$ meV (the
latter corresponds to the saturation value of Mn spin magnetization
$M$ for $x=0.05$). Crosses show spin and orbital contribution to
$M_c$.}
 \label{fig:mcvsp_6}
\end{figure}

\begin{figure}
\includegraphics*[width=90mm]{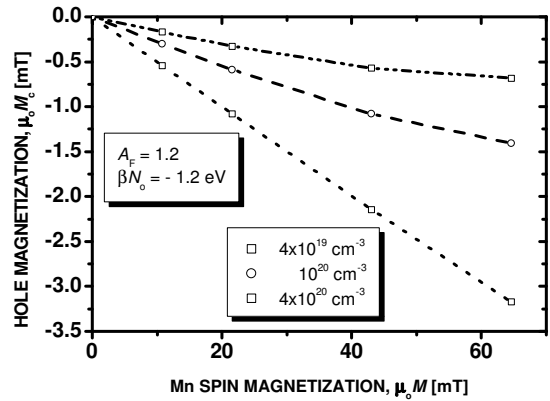}
\caption[]{Computed hole magnetization $M_c$ as a function of Mn
spin magnetization $M$ for various hole concentrations in
Ga$_{1-x}$Mn$_{x}$As.}
 \label{fig:mcvsm_7}
\end{figure}

The results of the computations are shown in
Fig.~\ref{fig:mcvsp_6}, where $M_c$ is plotted as a function of $p$
for $B_G = A_F\beta M/6g\mu_B = -30$ meV. In this paper, for sake
of comparison with experimental results we depict magnetization in
the SI units according to $\mu_oM$[T]=$4\pi 10^{-4}M$[emu]. It is
seen the diamagnetic (orbital) contribution to $M_c$ is negative
and dominant. The spin term is positive, which for the
antiferromagnetic sign of the p-d exchange integral $\beta$ points
to the negative sign of the hole Land\'e factor $g_h$. A visible
decrease of the spin contribution for the large $p$ corresponds to
a cross-over to the free electron value $g=2.0$ occurring when
$|\varepsilon_F|$ approaches $\Delta_o$. In Fig.~\ref{fig:mcvsm_7},
$M_c$ is plotted versus $M$ for various $p$. It is seen that for
the employed values of the parameters, $M_c$ reaches only 5\% of
$\mu_oM = 65$ mT, which corresponds to the saturation value of Mn
magnetization for $x=0.05$. A rather weak magnitude of $M_c$
results from a partial compensation of the spin and orbital
contributions to $M_c$ as well as from smaller concentration and
spin of the holes in comparison to those of the Mn ions. We
conclude that delocalized or weakly localized holes give a minor
contribution to the total magnetization. Accordingly $M_c$ is
neglected when determining the direction of easy axes and the
magnitude of anisotropy fields.

In view of the on-going experiments\cite{Ohno99b} on electrical
spin injection from (Ga,Mn)As, an important question arises what is
the degree of hole spin polarization ${\mathcal{P}}$ as a function
of $p$ and $B_G$. Furthermore, ${\mathcal{P}}$ appears to control
the magnitude of the extraordinary Hall effect. It is, therefore,
interesting to determine conditions, under which the usual
assumption about the linear relation between ${\mathcal{P}}$ and
magnetization of the Mn spins $M$ is fulfilled.

The contribution of all four hole subbands to the Fermi cross
section visible in Fig.~\ref{fig:fs_2} indicates that the exchange
splitting is too small to lead to the total spin polarization for
$p=3.5\times 10^{20}$ cm$^{-3}$ and $x$ = 0.05 ($|B_G| = 30$ meV).
Furthermore, a destructive effect of the spin-orbit interaction on
the magnitude of ${\mathcal{P}}$ can be expected. In order to
evaluate ${\mathcal{P}}\equiv 2\langle s_M\rangle/p$ we note that
according to Eq.~(\ref{eq:pd}),
\begin{equation}
{\mathcal{P}} = \frac{2g\mu_B}{\beta p}\frac{\partial
F_c(M)}{\partial M}. \label{eq:pol}
\end{equation}

\begin{figure}
\includegraphics*[width=90mm]{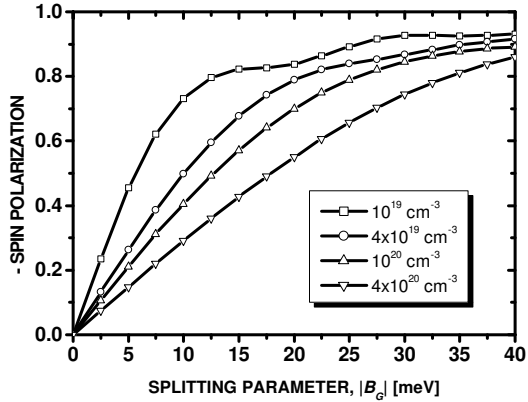}
\caption[]{Computed degree of spin polarization of the hole liquid
as a function of the spin splitting parameter for various hole
concentrations in  Ga$_{1-x}$Mn$_{x}$As ($B_G = -30$ meV
corresponds to the saturation value of Mn spin magnetization for
$x=0.05$). The polarization of the hole spins is oriented in the
opposite direction to the polarization of the Mn spins.}
 \label{fig:pola_8}
\end{figure}

Figure~\ref{fig:pola_8} presents the dependence of ${\mathcal{P}}$
on $B_G$ for the experimentally relevant range of $p$. We see that
$|{\mathcal{P}}|$ tends to saturate with $|B_G|$, and thus with
$M$. This means that for large values of the splitting $B_G$,
magnetization $M$ evaluated from the extraordinary Hall effect is
underestimated. At the same time, the calculation demonstrates that
despite the spin-orbit interaction $|{\mathcal{P}}|$ becomes
greater than 0.8 for $3|B_G| > |\varepsilon_F|$. This is due to the
fact that the redistribution of the holes over the spin subbands
occurs in the way which maximizes the gain of the exchange energy,
and thus the magnitude of $|{\mathcal{P}}|$. This in contrast to
the case $|\varepsilon_F| \gg |B_G|$, for which the spin
polarization is reduced by a factor greater than two from the value
corresponding to the absence of the spin-orbit coupling at low hole
concentrations. We also note that because of the antiferromagnetic
character of the p-d coupling ($\beta N_o<0$), the polarization of
the hole spins is oriented in the opposite direction than the
polarization of the Mn spins.

\subsection{Easy axis and anisotropy field}

Already early studies of a ferromagnetic phase in (Ga,Mn)As
epilayers demonstrated the existence of substantial magnetic
anisotropy.\cite{Ohno96b} Magnetic anisotropy is usually associated
with the interaction between spin and orbital degrees of freedom of
the {\em magnetic} electrons. According to the model in question,
these electrons are in the d$^5$ configuration. For such a case the
orbital momentum $L = 0$, so that no effects stemming from the
spin-orbit coupling are to be expected.  To reconcile the model and
the experimental observations, we note that the interaction between
the localized spins is mediated by the holes, characterized by a
non-zero orbital momentum. An important aspect of the present model
is that it does take into account the anisotropy of the
carrier-mediated exchange interaction associated with the
spin-orbit coupling in the host material, an effect difficult to
include within the standard approach to the RKKY interaction.

\begin{figure}
\includegraphics*[width=90mm]{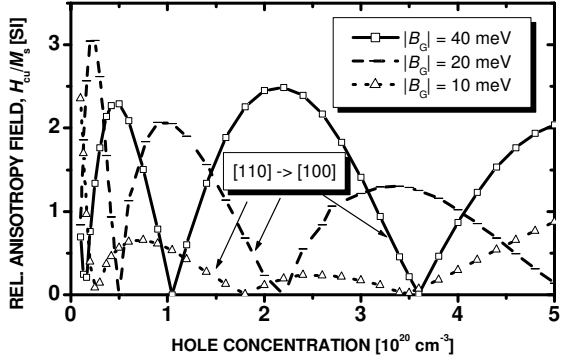}
\caption[]{Computed minimum magnetic field $H_{cu}$ (divided by
$M$) necessary to align magnetization $M$ along the hard axis for
cubic (unstrained) Ga$_{1-x}$Mn$_{x}$As film. As a function of the
hole concentration and the spin-splitting parameter $B_G$, the easy
and hard axes fluctuate alternatively between [110] and [100] (or
equivalent) directions in the plane of the film. The symbol [110]
$\rightarrow$ [100] means that the easy axis is along [110], so
that $H_{cu}$ is applied along [100] ($B_G = -30$ meV corresponds
to the saturation value of $M$ for Ga$_{0.95}$Mn$_{0.05}$As).}
 \label{fig:bcu_9}
\end{figure}

We start by considering an unstrained thin film with the [001]
crystal direction perpendicular to its plane. The linear response
is isotropic in cubic systems but magnetic anisotropy develops for
non-zero magnetization: the hole energy depends on the orientation
of $M$ in respect to crystal axes and, because of stray field
energy $E_d$, on the angle $\Theta$ between  $M$ and the normal to
the film surface. A computation of the hole energies $E_c[M]$ for
relevant parameters and [100], [110], and [111] orientations of
${\bm M}$ indicates the that $E_c[M]$ can be described by the
lowest order cubic anisotropy, so that, taking  the stray field
energy into account,\cite{Hube98}
\begin{eqnarray}
\nonumber
E_c(M,\Theta,\varphi) - E_c(M,\pi/2,0) =
K_d(M)\cos^2\Theta + \\
+ K_{c1}(M)(\sin^4\Theta\sin^2\varphi
\cos^2\varphi +\sin^2\Theta\cos^2\Theta),
\end{eqnarray}
where $K_d(M) =2\pi M^2$. For  $4K_d < -K_{c1}$ the easy axis is
oriented along [111] or equivalent directions. Otherwise, as is
usually the case for parameters of (Ga,Mn)As, ${\bm M}$  lies in
the (001) plane, and the easy axis is directed along  [100] for
$K_{c1}>0$ or along [110] (or equivalent) crystal axis in the
opposite case. It turns out that the sign of $K_{c1}$ depends on
the degree of occupation the hole subbands as well as on their
mixing by the p-d and $kp$ interactions. As a result, the easy axis
fluctuates between [100] and [110] as a function of $p$. To
quantify the strength of the cubic anisotropy in (Ga,Mn)As we
computed $K_{c1}(M)$, and then the minimum magnitude of an external
magnetic field $H_{cu} = 2|K_{c1}/M|$ (or $\mu_oH_{cu} =
2|K_{c1}/M|$ in the SI units), which aligns the spontaneous
magnetization ${\bm M}$ along the hard direction.
Figure~\ref{fig:bcu_9} shows how $H_{cu}/M$ and the direction of
the easy axis oscillates as a function of $p$ for various $B_G =
A_F\beta M/6g\mu_B$. As expected, $H_{cu}$ tends to zero when $B_G$
decreases. Nevertheless, for $|B_G| = 40$ meV ($\mu_oM \approx 85$
mT), $\mu_oH_{cu}$ up to 0.2 T can be expected. Since, however, the
orientation of the easy axis changes rapidly with $p$ and $B_G$,
intrinsic and extrinsic disorder--which leads to broadening of hole
subbands--will presumably diminish the actual magnitude of magnetic
anisotropy.

As shown in Fig.~\ref{fig:fs_2}, strain has a rather strong
influence on the valence subbands. It can, therefore, be expected
that magnetic properties resulting from the hole-mediated exchange
can be efficiently controlled by strain engineering. Indeed,
sizable lattice-mismatch driven strain is known to exist in
semiconductor layers. In some cases, particularly if epitaxy occurs
at appropriately low temperatures, such strain can persists even
beyond the critical thickness due to relatively high barriers for
the formation of misfit dislocations. We evaluate the magnitude of
resulting effects by using the Bir-Pikus hamiltonian,\cite{Bir74}
adopted for biaxial strain, as shown in Appendix A. Three
parameters control strain phenomena in the valence band: the
deformation potential $b$, taken as $b = -1.7$ eV,\cite{Bimb82} the
ratio of elastic moduli $c_{12}/c_{11}=0.453$,\cite{Bimb82} and the
difference between the lattice parameters of the substrate and the
layer, $\Delta a$. The latter is related to relevant components of
the strain tensor according to,
\begin{eqnarray}
 \epsilon_{xx} = \epsilon_{yy}= \Delta a/a;\\
 \epsilon_{zz} =
-2\epsilon_{xx}c_{12}/c_{11}.
\end{eqnarray}

\begin{figure}
\includegraphics*[width=90mm]{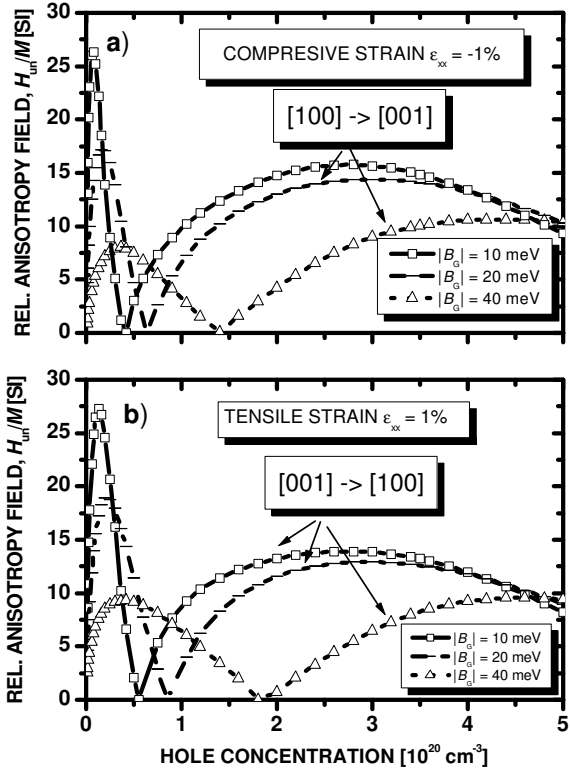}
\caption[]{Computed minimum magnetic field $H_{un}$ (divided by
$M$) necessary to align magnetization $M$ along the hard axis for
compressive (a) and tensile (b) biaxial strain in
Ga$_{1-x}$Mn$_{x}$As film for various values of the spin-splitting
parameter $B_G$. The easy axis is along [001] direction and in the
(001) plane at low and high hole concentrations for compressive
strain, respectively (a). The opposite behavior is observed for
tensile strain (b). The symbol [100] $\rightarrow$ [001] means that
the easy axis is along [100], so that $H_{un}$ is applied along
[001] ($B_G = -30$ meV corresponds to the saturation value of $M$
for Ga$_{0.95}$Mn$_{0.05}$As).}
 \label{fig:bun_10}
\end{figure}

\begin{figure}
\includegraphics*[width=90mm]{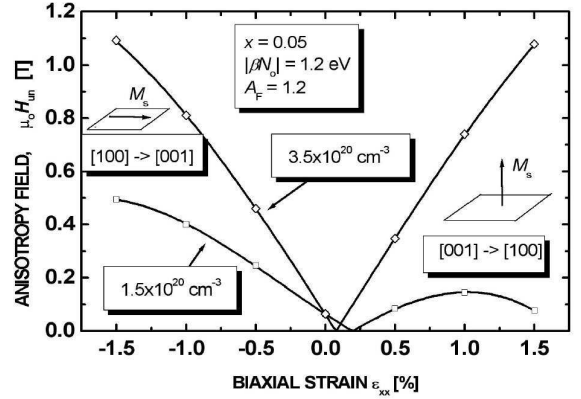}
\caption[]{Computed minimum magnetic field $H_{un}$ necessary to
align the saturation value of magnetization $M_s$ along the hard
axis as a function of biaxial strain component $\epsilon_{xx}$ for
two values of the hole concentrations in Ga$_{0.95}$Mn$_{0.05}$As.
The symbol [100] $\rightarrow$ [001] means that the easy axis is
along [100], so that $H_{un}$ is applied along [001].}
\label{fig:bunvse_11}
\end{figure}

We have found that biaxial strain has a rather small influence on
$T_C$. In the experimentally relevant range of hole concentrations
$5\times 10^{20} > p > 10^{20}$ cm$^{-3}$,  both tensile and
compressive strain diminish $T_C$. The relative effect attains a
maximum at $p \approx 2\times 10^{20}$ cm$^{-3}$, where
$[T_C(\epsilon_{xx)}-T_C(0)]/T_C(0) \approx -2.4$\% and $-4.9$\%
for $\epsilon_{xx} = 1$\% and $-1$\%, respectively. However, such a
strain leads to uniaxial anisotropy, whose magnitude can be much
greater than that resulting from either cubic anisotropy or stray
fields. The corresponding anisotropy field $H_{un}$ assumes the
form,
\begin{eqnarray}
H_{un}  = |2[E_c([001]) - E_c([100])]/M + 4\pi M| \mbox{[emu]};\\
 \mu_oH_{un}  =
|2[E_c([001]) - E_c([100])]/M + \mu_oM|\mbox{[SI]},
\end{eqnarray}
where the last term describes the stray-field effect. According to
results of computations presented in Fig.~\ref{fig:bun_10}, in the
region of such low hole concentrations that minority spin subbands
are depopulated, the easy axis takes the [001] direction for the
compressive strain, while is in the (001) plane for the opposite
strain. Such a behavior of magnetic anisotropy was recently noted
within a 4x4 model of the valence band.\cite{Abol00} However, for
the experimentally relevant hole concentrations and values of $B_G$
(Figs.~\ref{fig:bun_10} and \ref{fig:bunvse_11}) the easy axis is
oriented along [001] direction for the tensile strain, whereas
resides in the (001) plane for the case of unstrained or
compressively strained films. This important result, announced
already in our previous work,\cite{Diet00} is corroborated by the
experimental study,\cite{Ohno96b} in which either (Ga,In)As or GaAs
substrate was employed to impose tensile or compressive strain,
respectively. In particular, for the Ga$_{0.0965}$Mn$_{0.035}$As
film on GaAs, for which $\epsilon_{xx} = -0.2$\%, the anisotropy
field $\mu_oH_{un} = 0.4 \pm 0.1$ T is observed,\cite{Ohno96b} in
quantitative agreement with the theoretical results of
Fig.~\ref{fig:bunvse_11}.

Finally, we mention that strong strain effects may suggest the
importance of magnetostriction in the studied compounds. We have
not explored this issue yet, and note that prior to its examination
the question concerning the collective Jahn-Teller effect in
heavily doped p-type zinc-blende semiconductors has to be
addressed.

\subsection{Optical absorption and magnetic circular dichroism}
\label{sec:mcd}
 In the case of II-VI diluted magnetic semiconductors, detail
information on the exchange-induced spin-splitting of the bands,
and thus on the coupling between the effective mass electrons and
the localized spins has been obtained from magnetooptical
studies.\cite{Diet94,Furd88b} Similar
investigations\cite{Szcz99a,Besc99,Szcz96,Ando98} of (Ga,Mn)As led
to a number of surprises. The most striking was the opposite order
of the absorption edges corresponding to the two circular photon
polarizations in (Ga,Mn)As comparing to II-VI materials. This
behavior of circular magnetic dichroism (MCD) suggested the
opposite order of the exchange-split spin subbands, and thus a
different origin of the sp-d interaction in these two families of
DMS. A new light on the issue was shed by studies of
photoluminescence (PL) and its excitation spectra (PLE) in p-type
(Cd,Mn)Te quantum wells.\cite{Haur97,Koss00,Cibe98}  It has been
demonstrated that the reversal of the order of PLE edges
corresponding to the two circular polarizations results from the
Moss-Burstein effect, that is from the shifts of the absorption
edges associated with the empty portion of the valence subbands in
the p-type material. This model was subsequently applied to
interpret qualitatively the magnetoabsorption data for metallic
(Ga,Mn)As.\cite{Szcz99a}  Surprisingly, however, the anomalous sign
of the MCD was present also in non-metallic (Ga,Mn)As, in which EPR
signal from occupied Mn acceptors was seen.\cite{Szcz99b} Another
striking property of the MCD is a different temperature dependence
of the normalized MCD at low and high photon energies in
ferromagnetic (Ga,Mn)As.\cite{Besc99} This observation was taken as
an evidence for the presence of two spectrally distinct
contributions to optical absorption.\cite{Besc99}

We begin by noting that according to our two-fluid model, the
co-existence of strongly and weakly localized holes is actually
expected on the both sides of the MIT. Since the Moss-Burstein
effect operates for interband optical transitions involving weakly
localized states, it leads to the sign reversal of the MCD, also on
the insulating side of the MIT. Conversely, the existence of the
MCD sign reversal can be taken as an experimental evidence for the
presence of the Fermi liquid--like states on the insulating side of
the MIT.

\begin{figure}
\includegraphics*[width=90mm]{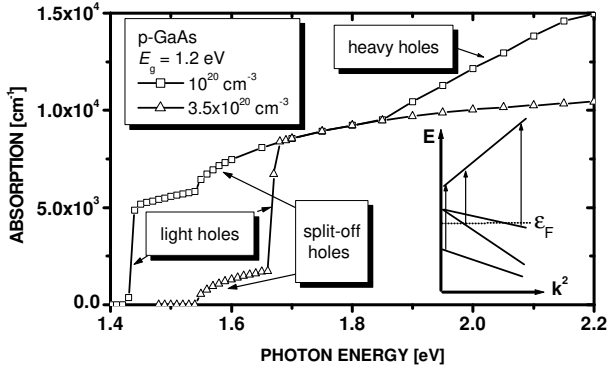}
\caption[]{Absorption edge in p-GaAs computed for two values of the
hole concentrations. Inset shows three kinds of possible
photon-induced transitions corresponding to particular valence
subbands. Absorption edges associated with electron transitions
from particular subbands are labeled.} \label{fig:abs0_12}
\end{figure}

In order to shed some light on those issues we calculate absorption
and MCD spectra in a model that takes the complex structure of the
valence band into account. The band-gap $E_g$ is expected to depend
on both Mn and hole concentration due to the alloy and band
narrowing effects.  To take this dependence into account as well as
to include disorder-induced band tail effects\cite{Case76} we
assume, guided by experimental results to be discussed below, that
$E_g=1.2$ eV. Our computation of the absorption coefficient
$\alpha(\omega)$ are performed according to a scheme outlined in
Appendix B, taking the electron effective mass $m_e^*=0.07m_o$, the
Kane momentum matrix element $P=9.9\times 10^{-8}$ eVcm, and the
refractive index $n_r =3.5$. As shown in Fig.~\ref{fig:abs0_12},
contributions to $\alpha$ originating from particular valence bands
are clearly visible. Because of the Moss-Burstein shift, the onset
and the form of $\alpha(\omega)$ for particular transitions depend
on the hole concentration. In particular, $\alpha(\omega)$
corresponding to the light-hole band exhibits a step-like behavior
which, in the case of the heavy hole band, is broadened by warping.
While no quantitative data on $\alpha(\omega)$ are available for
(Ga,Mn)As, the computed magnitude and spectral dependence of
$\alpha(\omega)$ reproduce correctly experimental results for
p-GaAs.\cite{Case75}

\begin{figure}
\includegraphics*[width=90mm]{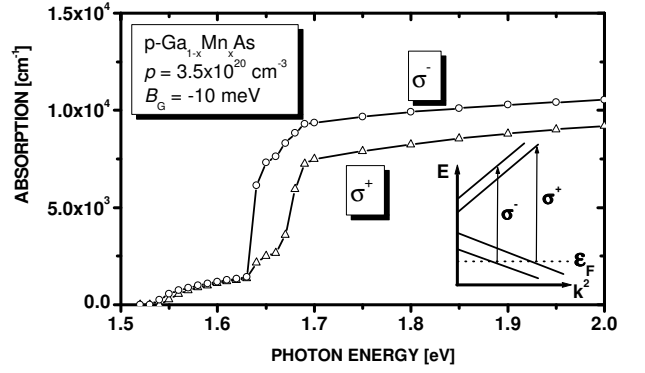}
\caption[]{Absorption edge for two circular polarizations in
p-(Ga,Mn)As computed for spin-splitting parameter $B_G=-10$ meV and
hole concentration $3.5\times 10^{20}$ cm$^{-3}$. Inset shows how
the Fermi see of the holes reverses relative positions of the edges
corresponding to $\sigma^+$ and $\sigma^-$ polarizations. In
general, electron transitions from six valence subbands contribute
to optical absorption.}
 \label{fig:absbg_13}
\end{figure}

The influence of the sp-d band splittings on the absorption edge is
shown in Fig.~\ref{fig:absbg_13}. The computation are carried out
for the Faraday configuration and with the value of the s-d
exchange energy $\alpha N_o = 0.2$ eV observed in II-VI
semiconductors. The theoretical results confirm that the
Moss-Burstein effect accounts for the sign reversal of the magnetic
circular dichroism (MCD). The energy splitting of the absorption
edge depends on $\omega$ but its magnitude is similar to that
observed experimentally.\cite{Szcz99a} A detail comparison requires
from one hand experimental information on the absolute values of
$\alpha(\omega)$ and, on the other, more careful consideration of
band tailing effects. Furthermore, contributions from
intra-valence-band and from intra-d-shell transitions are expected
at low and high energy wings of the absorption edge, respectively.
We predict that not only the former but also the latter are
substantially enlarged in p-type materials. Indeed, the empty
valence band states allow for admixtures of p-like states to the
localized d orbitals.

The magnetization-induced splitting of the bands is seen to lead to
a large energy difference between the positions of the absorption
edges corresponding to the two opposite circular polarizations.
This may cause an unusual dependence of the low-energy onset of MCD
on magnetization, and thus on temperature. In particular, a
standard assumption about the proportionality of MCD and
magnetization becomes invalid. To find out whether the latter is
responsible for the anomalous temperature dependence of MCD at low
photon energies,\cite{Besc99} we  compute the differential
transmission coefficient that was examined
experimentally\cite{Besc99}
\begin{equation}
\Delta = (T^+-T^-)/(T^++T^-).
\end{equation}
Here $T^+$ and $T^-$ are the transmission coefficients for right
and left circularly polarized light. To take the effect of
interference into account,\cite{Besc99} these coefficients are
calculated for the actual layout of the samples, which consisted of
a transparent (Ga,Al)As etching stop layer and the absorbing
(Ga,Mn)As film, each 200 nm thick. The same value of the refractive
index $n_r=3.5$ are adopted for both compounds.

\begin{figure}
\includegraphics*[width=90mm]{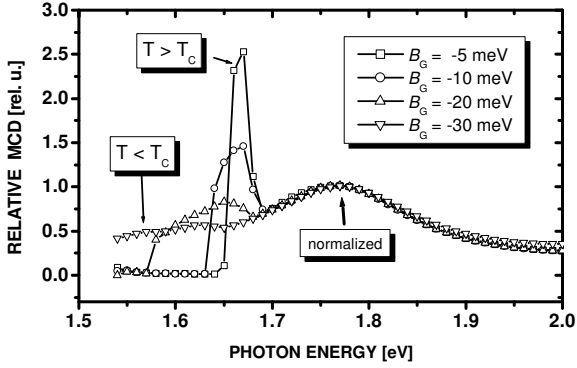}
\caption[]{Spectral dependence of magnetic circular dichroism in
p-(Ga,Mn)As computed for hole concentration $3.5\times 10^{20}$
cm$^{-3}$ and various spin-splitting parameters $B_G$. The
magnitudes of MCD at given $B_G$ are normalized by its value at
1.78 eV.}
 \label{fig:mcd_14}
\end{figure}

Figure \ref{fig:mcd_14} shows the ratio
$\Delta(\omega)/\Delta$(1.78 eV) computed for $p=3.5\times 10^{20}$
cm$^{-3}$ and various $B_G \sim M$. In the range of high photon
energies, $\omega >1.6$ eV, the results collapse into one curve for
all values of $B_G$ . This means that
$\Delta(\omega)=M(T)f(\omega)$ in this range, where $f(\omega) \sim
T^{-1}\partial T/\partial\omega$. However, in the region of the
absorption edge, the dependence of $\Delta$ on $B_G$ is by no means
linear, so that the normalized values $\Delta(\omega)/\Delta$(1.78
eV) do not follow any single curve. As seen
$\Delta(\omega)/\Delta$(1.78 eV) peaks at the greatest value the
smallest $B_G$. This is the behavior found
experimentally.\cite{Besc99} We conclude that the two observed
distinct spectroscopic regions\cite{Besc99} correspond to standard
band to band transitions, for which the proportionality of $\Delta$
to $B_G$ holds, and to the onset of the absorption edge that is
shifted and make more steep by the Moss-Burstein effect. Actually,
the peak values of $\Delta(\omega)/\Delta$(1.78 eV) determined
numerically for the low energy region are even greater than that
observed in (Ga,Mn)As,\cite{Besc99} presumably because of
scattering broadening of the absorption edge, neglected in our
model.

\section{CHEMICAL TRENDS}
\subsection{Material parameters}

The ability of the present model to describe successfully various
aspects of the ferromagnetism in (Ga,Mn)As  as well as in
(Zn,Mn)Te,\cite{Ferr99,Diet00,Ferr00} has encouraged us to extend
the theory towards other p-type diluted magnetic semiconductors. In
this Section,  we present material parameters that have been
adopted for the computations presented elsewhere.\cite{Diet00} We
supplement also the previous results\cite{Diet00} by the data for
(In,Mn)N, (Cd,Mn)S, and (Cd,Mn)Se. For concretness, we will assume
that 5\% of the cation sites are occupied by the Mn ions in the 2+
charge state, and that the corresponding localized spins $S=5/2$
are coupled by the indirect exchange interaction mediated by
$3.5\times10^{20}$ holes per cm$^3$. The enhancement effect of the
exchange interaction among the holes in described by the Fermi
liquid parameter $A_F = 1.2$. As explained in Sec.~\ref{sec:two},
no influence of the antiferromagnetic superexchange is taken into
account in the case of the group III-V and IV semiconductors, in
which the Mn supplies both localized spins and holes. By contrast,
it is assumed that in the case of II-VI semiconductors $x_{eff} =
0.0297$ for $x=0.05$ and  $T_{AF}=1$ K, except for (Zn,Mn)Te, where
$T_{AF} = 2.9$ K,\cite{Shap86,Twar84a} and (Cd,Mn)Te, for which
$T_{AF} = 1.5$ K.\cite{Gaj79}

The values of the parameters that are used to determined chemical
trends are summarized in Tables \ref{table:zinc} and
 \ref{table:wurzite} for cubic and wurzite semiconductors,
respectively. If no experimentally determined values are available,
the effective mass parameters are determined by fitting the
appropriate $kp$ model to results of band structure computations.
No lattice polaron corrections are taken into consideration. Since
we are interested in a relatively small concentration of magnetic
ions, $x=0.05$, the effect of the Mn incorporation upon the lattice
and band structure parameters is disregarded.

In the case of ZnSe, the recently determined\cite{Lank96,Hols85}
values of the Luttinger parameters $\gamma_i$ lead to a negative
hole mass for the [110] crystallographic direction, an effect not
supported by the existing theoretical studies of the valence band
structure in this material.\cite{Vogl96,Blac99} Accordingly, an
older set\cite{Veng79}  of $\gamma_i$ had been taken for the
previous calculation.\cite{Diet00} The present values of $\gamma_i$
are within experimental uncertainties of the current
determinations\cite{Lank96} and, at the same time, lead to a good
description of the computed band structure of the valence band in
ZnSe.\cite{Vogl96}

\begin{figure}
\includegraphics*[width=90mm]{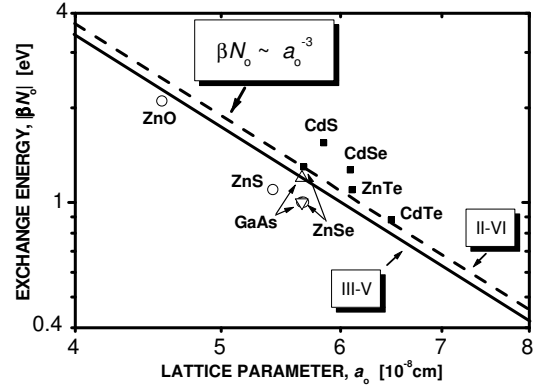}
\caption[]{Energy of p-d exchange interaction for various materials
containing 5\% of Mn as a function of lattice parameter. The values
shown by solid squares were determined from excitonic splittings in
the magnetic field,\cite{Beno92,Twar84a,Gaj79,Twar84b} while the
empty symbols denote values evaluated from photoemission
data.\cite{Okab98,Mizo99} Solid and dashed lines represent formulae
adopted for the determination of the exchange energy for other
materials, as shown in Tables \ref{table:zinc} and
\ref{table:wurzite}.} \label{fig:beta_15}
\end{figure}

In addition to the spin density of states at the Fermi level, the
Curie temperature is proportional to the square of the p-d exchange
integral $\beta$.  Figure \ref{fig:beta_15} presents the magnitudes
of the exchange energy $\beta N_o$ as determined by photoemission
and magnetooptical studies for various DMS containing about 5\% on
Mn. The values of $|\beta N_o|$ are seen to increase when the
lattice parameter decreases. This trend stems\cite{Mizo99} from the
corresponding changes in the charge transfer and correlation
energies as well as from a dependence of the p-d hybridization
energy on the bond length $b$.\cite{Mizo99} It should, however, be
recalled that $b$, in contrast to the average value of the lattice
constant, does not  obey the Vegard low in alloys but rather
conserves the value corresponding to the end
compounds.\cite{Balz85}

Thus, in order to obtain the values of  $\beta N_o$ for materials,
for which no direct determination is available, guided by the
results presented in Fig.~\ref{fig:beta_15},  we assume $\beta N_o
\sim a_o^{-3}$, {\it i.\,e.},  $\beta$ = const. More explicitly,
for group III-V and IV semiconductors we take
\begin{equation}
\label{eq:III-V} \beta_{III-V} = \beta(\mbox{GaMnAs}).
\end{equation}
Similarly, for the II-VI materials
\begin{equation}
\label{eq:II-VI} \beta_{II-VI} = \beta(\mbox{ZnMnSe}),
\end{equation}
where the p-d energy $\beta N_o{\mbox{(GaMnAs}}) = -1.$2
eV,\cite{Okab98} and $\beta N_o{\mbox{(ZnMnSe)}} = -1.3$
eV.\cite{Twar84b}

\subsection{Curie temperatures}

\begin{figure}
\includegraphics*[width=90mm]{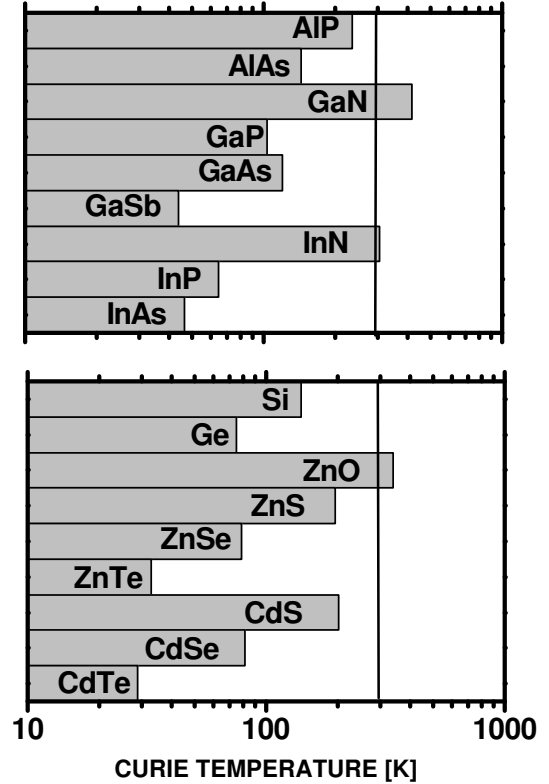}
\caption[]{Curie temperatures evaluated for various III-V (a) as
well group IV and II-VI semiconducting compounds (b) containing 5\%
of Mn in 2+ charge state and $3.5\times 10^{20}$ holes per
cm$^{3}$. Material parameters adopted for the calculation are
displayed in Tables \ref{table:zinc} and \ref{table:wurzite}.}
\label{fig:tcall_16}
\end{figure}

Figure \ref{fig:tcall_16} presents the calculated values of the
Curie temperature $T_C$ for III-V and II-VI semiconductors
containing 5\% of Mn and $3.5\times 10^{20}$ holes per cm$^{3}$.
The data for Si and Ge are also included. The most remarkable
result is a strong increase of $T_C$ for materials consisting of
lighter elements. Actually, $T_C$ exceeding room temperature is
expected for GaN, InN, and ZnO for the assumed values of the Mn and
hole concentrations. It has been checked for GaN that the value of
$T_C$ for the zinc-blend modification of this material is by 6\%
greater than that for wurzite case.

By comparing results of numerical calculations with the general
Eq.~(\ref{eq:tc}) for $T_C$ three interrelated mechanisms
accounting for the chemical trends visible in
Fig.~\ref{fig:tcall_16} can be identified. First, the reduction of
the spin density-of-states, and thus $T_C$ by the spin-orbit
interaction ceases to operate in materials with light anions.
Second, the effective mass, and thus the density-of-states tend to
increase for materials with stronger bonds. Finally, the smaller
lattice constant at given $x$ corresponds to the greater value of
$N_ox$, density of the magnetic ions. It should be noted at this
point that $T_C$ is proportional to $\beta^2$, assumed here to be
material independent. This assumption corresponds, however, to a
strong increase of $|\beta N_o|$ with decreasing lattice constant,
as shown in Fig.~\ref{fig:beta_15}.

It can be expected that the chemical trends established here are
not altered by the uncertainties in the values of the relevant
parameters.  Our evaluations of the strength of the ferromagnetic
interactions mediated by the holes is qualitatively valid for Mn,
as well as for other magnetic ions, provided that two conditions
are met. First, the magnetic electrons stay localized and do not
contribute directly to the Fermi sphere. Secondly, the holes are
delocalized and, in particular, do not form small magnetic
polarons, such as Zhang-Rice singlets.\cite{Zhan88}

\section{Summary and outlook}

In this paper, theory of ferromagnetism in p-type zinc-blende and
wurzite semiconductors containing a sizable concentration of
magnetic ions has been proposed. In has been argued that over the
relevant range of the hole densities the ferromagnetic coupling
between the localized spins is primarily mediated by delocalized or
weakly localized holes residing in p-like valence band.
Accordingly, particular attention has been paid to incorporate into
the Zener model effects of $kp$ and spin-orbit interactions as well
as of biaxial strain. It has been demonstrated that theory
describes qualitatively, and often quantitatively, a body of
experimental results accumulated over the recent years for
(Ga,Mn)As. In particular, Curie temperature, saturation value of
magnetization, hole spin magnetization and polarization, magnetic
anisotropies and magnetoelastic effects, optical absorption and
magnetic circular dichroism have been interpreted.

Giant negative magnetoresistance,  sharp field-induced
insulator-to-metal transition,\cite{Wojt86} and a sizable increase
of high frequency conductivity with the magnetic field\cite{Chmi87}
were observed in p-(Hg,Mn)Te. Those findings were attributed to the
growing participation of the light holes in transport when  the p-d
exchange splitting increases.\cite{Wojt86,Chmi87} This implies a
shift of the Drude weight from high to low frequencies as a
function of the valence-band splitting. Such effects, in both
d.c.\cite{Oiwa97,Kats98} and a.c.\cite{Hira99,Naga99} conductivity
have more recently been detected in (Ga,Mn)As,  and can
qualitatively be interpreted in the same way. This provides an
additional support for our conclusion about the similarity of the
mechanisms accounting for the hole-mediated exchange interaction in
II-VI and III-V magnetic semiconductors. However, our work
identifies also an aspect of ferromagnetism, which points to a
difference between those two families of magnetic semiconductors.
In the case of II-VI compounds, a short-range antiferromagnetic
superexchage lowers the magnitude of $T_C$. This lowering appears
to be much less efficient in III-V semiconductors, where the Mn
ions act as acceptors, compensated partly by donor defects. Thus,
the localized holes reside preferentially on the Mn pairs, so that
the hole-mediated ferromagnetic coupling (a variant of Zener's
double exchange) can overcompensate the antiferromagnetic
superexchange.

The model put forward here has also been used to explore the
expected chemical trends. It has been found that particularly large
value of the Curie temperature are expected for materials built up
from light elements. Important issues of solubility limits and
self-compensation need, however, to be addressed experimentally. In
particular, the pinning of the Fermi energy by AX-type centers or
other defects can preclude the increase of the hole concentration
in many systems. High pressure research can shed some light on this
issue. Since, in general, III-V compounds can easier be doped by
impurities that are electrically active, whereas a large quantity
of transition metals can be incorporated into II-VI materials, a
suggestion has been put forward to grow magnetic III-V/II-VI short
period superlattices.\cite{Kama99} Further numerical and
experimental studies of magnetic semiconductors as well as of
heavily p-doped non-magnetic systems are expected to improve our
understanding of the hole-mediated ferromagnetism in zinc-blende
and wurzite compounds. This, together with exploration of novel
quantum structures as well as of co-doping and co-alloying, may
lead to fabrication of functional systems.

On the theoretical side, further work is necessary to evaluate
quantitative corrections to the mean field theory brought about by
thermodynamic fluctuations of magnetization.  Effects of disorder
associated with both random distribution of magnetic ions and
fluctuations of carrier density near the metal-insulator transition
are other open issues. In particular, unknown is the nature of
evolution of static and dynamic magnetic phenomena on approaching
the strongly localized regime. The above list of interesting
problems is by no means exhausting. With no doubt we will soon
witness many unforeseen developments in the field of
carrier-mediated ferromagnetism in semiconductors.

\section*{Acknowledgments}
The work at Tohoku University was supported by the Japan Society
for the Promotion of Science and by the Ministry of Education,
Japan; the work in Poland by State Committee for Scientific
Research, Grant No. 2-P03B-02417, and by Foundation for Polish
Science.

\widetext
\appendix
\section{Effective-mass hamiltonians}

The purpose of this Appendix is to provide the explicit form of the
effective mass hamiltonian that, in addition to the standard $kp$
and strain terms,\cite{Bir74} contains a contribution of the p-d
exchange interaction in the molecular field approximation.  The
latter constitutes a generalization of the previous
approaches\cite{Gaj78,Furd88b} by allowing for the arbitrary
orientation of the magnetization $M$ with respect to the crystal
axes.

Zinc-blende semiconductors are considered first. We take explicitly
into account four $\Gamma_8$ and two $\Gamma_7$ valence subbands,
for which we chose the basis functions in the form:
\begin{eqnarray}
u_1 &=& \frac{1}{\sqrt{2}}(X+\mbox{i}Y)\uparrow,\\
 u_2&=&\mbox{i}\frac{1}{\sqrt{6}}[(X+\mbox{i}Y)\downarrow -
 2Z\uparrow],\\
 u_3&=&\frac{1}{\sqrt{6}}[(X-\mbox{i}Y)\uparrow +
 2Z\downarrow],\\
u_4 &=& \mbox{i}\frac{1}{\sqrt{2}}(X-\mbox{i}Y)\downarrow,\\
u_5&=&\frac{1}{\sqrt{3}}[(X+\mbox{i}Y)\downarrow +Z\uparrow],\\
u_6&=&\mbox{i}\frac{1}{\sqrt{3}}[-(X-\mbox{i}Y)\uparrow +
 Z\downarrow],
\end{eqnarray}

where $X,Y,$ and $Z$ denote Kohn-Luttinger amplitudes, which for
the symmetry operations of the crystal point group transform as
$p_x, p_y,$ and $p_z$ wave functions of the hydrogen atom.

 In the above basis the corresponding Luttinger-Kohn
matrices assume the form:

\subsection{$kp$  matrix}
\begin{equation}
H_{kp} = -\frac{\hbar^2}{2m_o} \left[
\begin{array}{cccccc}
 P+Q&L&M&0&\mbox{i}L/\sqrt{2}&-\mbox{i}\sqrt{2}M\\
 L^*&P-Q&0&M&-\mbox{i}\sqrt{2}Q&\mbox{i}\sqrt{3/2}L\\
 M^*&0&P-Q&-L&-\mbox{i}\sqrt{3/2}L^*&-\mbox{i}\sqrt{2}Q\\
 0&M^*&-L^*&P+Q&-\mbox{i}M^*&-\mbox{i}L^*/\sqrt{2}\\
 -\mbox{i}L^*/\sqrt{2}&\mbox{i}\sqrt{2}Q&\mbox{i}\sqrt{3/2}L&
 \mbox{i}\sqrt{2}M&P+\Delta&0\\
\mbox{i}\sqrt{2}M^*&-\mbox{i}\sqrt{3/2}L^*&\mbox{i}\sqrt{2}Q&
\mbox{i}L/\sqrt{2}&0&P+\Delta
\end{array}\right].
\end{equation}
Here,
\begin{eqnarray}
P&=& \gamma_1k^2,\\
 Q&=&\gamma_2(k_x^2 + k_y^2 - 2k_z^2),\\
 L&=&-\mbox{i}2\sqrt{3}\gamma_3(k_x - \mbox{i}k_y)k_z,\\
 M&=&\sqrt{3}[\gamma_2(k_x^2 - k_y^2) -\mbox{i}2\gamma_3k_xk_y],\\
 \Delta&=&2m_o\Delta_o/\hbar^2.
\end{eqnarray}

\subsection{p-d exchange matrix}
\begin{equation}
H_{pd} = B_G\left[\begin{array}{cccccc}
 3b_xw_z&\mbox{i}\sqrt{3}b_xw_-& 0& 0& \sqrt{6}b_xw_-&  0\\
  -\mbox{i}\sqrt{3}b_xw_+& (2b_z-b_x)w_z& 2\mbox{i}b_zw_-& 0&
  \mbox{i}\sqrt{2}(b_x+b_z)w_z& -\sqrt{2}b_zw_-\\
   0& -2\mbox{i}b_zw_+& -(2b_z-b_x)w_z& \mbox{i}\sqrt{3}b_xw_-&
     \sqrt{2}b_zw_+& -\mbox{i}\sqrt{2}(b_x+b_z)w_z\\
  0& 0&-\mbox{i}\sqrt{3}b_xw_+&-3b_xw_z& 0&-\sqrt{6}b_xw_+\\
       \sqrt{6}b_xw_+&-\mbox{i}\sqrt{2}(b_x+b_z)w_z&
        \sqrt{2}b_zw_-&  0&-(2b_x-b_z)w_z& \mbox{i}b_zw_-\\
  0& -\sqrt{2}b_zw_+& \mbox{i}\sqrt{2}(b_x+b_z)w_z&
   -\sqrt{6}b_xw_-& -\mbox{i}b_zw_+&(2b_x-b_z)w_z
\end{array}\right].
\end{equation}
Here,
\begin{eqnarray}
B_G&=&\beta M/6g\mu_B,\\
 w_z&=&M_z/M,\\
 w_{\pm}&=&(M_x\pm\mbox{i}M_y)/M,\\
 b_z&=&\beta_z/\beta,\\
 b_x&=&\beta_x/\beta,
\end{eqnarray}
where in cubic materials $b_z=b_x =1$.

\subsection{Biaxial strain matrix}
\begin{equation}
H_{bs} = b\left[\begin{array}{cccccc}
 -Q_{\epsilon}& 0& R_{\epsilon}&0&0&-\mbox{i}\sqrt{2}R_{\epsilon}\\
 0&Q_{\epsilon}&0&R_{\epsilon}&\mbox{i}\sqrt{2}Q_{\epsilon}&0\\
 R_{\epsilon}&0&Q_{\epsilon}&0&0&\mbox{i}\sqrt{2}Q_{\epsilon}\\
 0&R_{\epsilon}&0&-Q_{\epsilon}&-\mbox{i}\sqrt{2}R_{\epsilon}&0\\
 0&-\mbox{i}\sqrt{2}Q_{\epsilon}&0&\mbox{i}\sqrt{2}R_{\epsilon}&0&0\\
 \mbox{i}\sqrt{2}R_{\epsilon}&0&-\mbox{i}\sqrt{2}Q_{\epsilon}&0&0&0
\end{array}\right].
\end{equation}
Here, $b$ is the deformation potential and
\begin{eqnarray}
 Q_{\epsilon} &=& \epsilon_{zz}  - ( \epsilon_{xx}+ \epsilon_{yy})/2,\\
 R_{\epsilon} &=& \sqrt{3}(\epsilon_{xx}  - \epsilon_{yy})/2.
\end{eqnarray}

Since we are interested in the effect of the biaxial strain in the
(001) plane, only the terms involving the diagonal components
$\epsilon_{ii}$ of the deformation tensor are included. For the
same reason, we allow for the corresponding anisotropy of the
exchange integrals $\beta_x = \beta_y \neq \beta_z$, though we
expect that to a good accuracy $\beta_x = \beta_z$ in real systems.
The latter is assumed in the main body of the paper.

In the presence of the magnetic field $B$ the Luttinger-Kohn $kp$
matrix is a sum of the Zeeman  and  Landau parts, $H_{kp}=H_Z+H_L$,
where $H_L$ is written below for ${\bm{B}}\parallel$ [001] and
neglecting some terms proportional to
$\gamma_2-\gamma_3$.\cite{Good61}

\begin{eqnarray}
H_{Z} = -g_o\mu_BB\left[\begin{array}{cccc}
 3\kappa w_z/2& \mbox{i}\sqrt{3}\kappa w_-/2& 0&...\\
 -\mbox{i}\sqrt{3}\kappa w_+/2& \kappa w_z/2&\mbox{i}\kappa w_-&...\\
   0 & -\mbox{i}\kappa w_+& -\kappa w_z/2&...\\
   0&0&-\mbox{i}\sqrt{3}\kappa w_+/2&...\\
   -\sqrt{6}(\kappa+1)w_+/4& \mbox{i}\sqrt{2}(\kappa+1)w_z/2&
    -\sqrt{2}(\kappa+1)w_-/4&...\\
   0& \sqrt{2}(\kappa+1)w_+/4& -\mbox{i}\sqrt{2}(\kappa+1)w_z/2&...
\nonumber \end{array} \right]
\\
 \left[
\begin{array}{cccc}
 ...&0&-\sqrt{6}(\kappa+1)w_-/4& 0\\
...&0& -\mbox{i}\sqrt{2}(\kappa+1)w_z/2& \sqrt{2}(\kappa+1)w_-/4\\
...&\mbox{i}\sqrt{3}\kappa
w_-/2&-\sqrt{2}(\kappa+1)w_+/4&\mbox{i}\sqrt{2}(\kappa+1)w_z/2\\
...&-3\kappa w_z/2& 0& \sqrt{6}(\kappa+1)w_+/4 \\
...&0&(\kappa+1/2)w_z& -\mbox{i}(\kappa +1/2)w_-\\
 ...&\sqrt{6}(\kappa+1)w_-/4&\mbox{i}(\kappa+1/2)w_+& -(\kappa+1/2)w_z
\end{array}
\right],
\end{eqnarray}
where $g_o$ is the Land\'e factor of the free electron, and
\begin{eqnarray}
w_z &=& B_z/B \\
 w_{\pm} &=& B_x \pm \mbox{i}B_y.
 \end{eqnarray}

 For the Landau quantum number $n>0$
\begin{eqnarray}
H_{L} = -\frac{\hbar^2}{2m_o}\left[\begin{array}{cccc}
 h_h+s(\gamma_1+\gamma_2)(2n-1)& b\sqrt{n}& c\sqrt{n(n+1)}&...\\
  -b\sqrt{n} h_l+s(\gamma_1-\gamma_2)(2n+1)&   0& c\sqrt{(n+1)(n+2)}&...\\
  c\sqrt{n(n+1)}& 0&   h_l+s(\gamma_1-\gamma_2)(2n+3)&...\\
   0&  c\sqrt{(n+1)(n+2)}& b\sqrt{n+2}&...\\
   \mbox{i}b\sqrt{n/2}& \mbox{i}q+\mbox{i}\sqrt{2}\gamma_2s(2n+1)&
    \mbox{i}b\sqrt{3(n+1)/2}&...\\
  \mbox{i}c\sqrt{2(n+1)n}& \mbox{i}b\sqrt{3(n+1)/2}&
  \mbox{i}q+\mbox{i}\sqrt{2}\gamma_2s(2n+3)&...
\end{array}
\right] \nonumber
\\
\left[
\begin{array}{cccc}
...&0& \mbox{i}b\sqrt{n/2}& -\mbox{i}c\sqrt{2n(n+1)}\\
 ...& -\mbox{i}q-\mbox{i}\sqrt{2}s\gamma_2(2n+1)&\mbox{i}b\sqrt{3(n+1)/2}\\
 ...& -b\sqrt(n+2)& \mbox{i}b\sqrt{3(n+1)/2}&
  -\mbox{i}q-\mbox{i}\sqrt{2}\gamma_2s(2n+3)\\
 ...& h_h+(\gamma_1+\gamma_2)s(2n+5)& -\mbox{i}c\sqrt{2(n+1)(n+2)}&
   \mbox{i}b\sqrt{(n+2)/2}\\
 ...& \mbox{i}c\sqrt{2(n+2)(n+1))}& h_s+\gamma_1s(2n+1)&0\\
 ...& \mbox{i}b\sqrt{(n+2)/2}&   0 &h_s+\gamma_1s(2n+3)
\end{array}
\right].
\end{eqnarray}
 Here,
\begin{eqnarray}
s&=&eB/\hbar,\\
 h_h &=& (\gamma_1-2\gamma_2)k_z^2,\\
      h_l &=& (\gamma_1+2\gamma_2)k_z^2,\\
      h_s &=& \gamma_1k_z^2,\\
      q&=&-2\sqrt{2}\gamma_2k_z^2,\\
      b&=&-\mbox{i}2\sqrt{6s}\gamma_3k_z;\\
      c&=&\sqrt{3}s(\gamma_2+\gamma_3).
\end{eqnarray}

 If $n=0$, the wave function basis does not
contain the $u_1$ term, so that
\begin{eqnarray}
H_{L} = -\frac{\hbar^2}{2m_o}\left[
\begin{array}{cccc}
 h_l+s(\gamma_1-\gamma_2)(2n+1)&   0&\sqrt{(n+2)(n+1)}&...\\
 0&  h_l+s(\gamma_1-\gamma_2)(2n+3)& -b\sqrt{n+2}&...\\
    c\sqrt{(n+2)(n+1)}& b\sqrt{n+2}&   h_h+(\gamma_1+\gamma_2)s(2n+5)&...\\
   \mbox{i}q+\mbox{i}\sqrt{2}\gamma_2s(2n+1)& \mbox{i}b\sqrt{3(n+1)/2}& \mbox{i}c\sqrt{2(n+2)(n+1)}&...\\
  \mbox{i}b\sqrt{3(n+1)/2}& \mbox{i}q+\mbox{i}\sqrt{2}\gamma_2s(2n+3)& \mbox{i}b\sqrt{(n+2)/2}&...\\
\end{array}
\right] \nonumber\\ \left[
\begin{array}{ccc}
 ...&-\mbox{i}q-\mbox{i}\sqrt{2}s\gamma_2(2n+1)&\mbox{i}b\sqrt{3(n+1)/2}\\
 ...&\mbox{i}b\sqrt{3(n+1)/2}& -\mbox{i}q-\mbox{i}\sqrt{2}\gamma_2s(2n+3)\\
 ...& -\mbox{i}c\sqrt{2(n+2)(n+1)}& \mbox{i}b\sqrt{(n+2)/2}\\
 ...&h_s+\gamma_1s(2n+1)&0\\
 ...&   0   &   h_s+\gamma_1s(2n+3)
\end{array}
\right].
\end{eqnarray}

If $n=-1$, the wave function basis does not contain the $u_1$,
$u_2$, and $u_5$ terms, so that
\begin{equation}
H_{L} = -\frac{\hbar^2}{2m_o}\left[\begin{array}{ccc}
 h_l+s(\gamma_1-\gamma_2)(2n+3)& -b\sqrt{n+2}&
 -\mbox{i}q-\mbox{i}\sqrt{2}\gamma_2s(2n+3)\\
     b\sqrt{n+2}&   h_h+(\gamma_1+\gamma_2)s(2n+5)&
     \mbox{i}b\sqrt{(n+2)/2}\\
 \mbox{i}q+\mbox{i}\sqrt{2}\gamma_2s(2n+3)&
 \mbox{i}b\sqrt{(n+2)/2}&h_s+\gamma_1s(2n+3)
\end{array}\right].
\end{equation}

If $n$=-2, the wave function basis contains only $u_4$ term, so
that
\begin{equation}
H_{L} = -\frac{\hbar^2}{2m_o}[h_h+(\gamma_1+\gamma_2)s].
\end{equation}

The solution of the eigenvalue and eigenfunction problem for the
hamiltonian $H=H_{kp} + H_{pd} + H_{bs}$ or, in the magnetic field,
$H=H_{L} + H_{Z} + H_{pd} + H_{bs}$ gives the hole energies
$\varepsilon_i$ and eigenvectors ${a_i^{(j)}}$ for the six relevant
hole subbands.

\subsection{Wurzite compounds}
For wurzite compounds, we chose the basis in the form,
\begin{eqnarray}
u_1 &=& -\frac{1}{\sqrt{2}}(X+\mbox{i}Y)\uparrow,\\
 u_2&=&\frac{1}{\sqrt{2}}(X-\mbox{i}Y)\uparrow,\\
 u_3&=&Z\uparrow,\\
u_4 &=& \frac{1}{\sqrt{2}}(X-\mbox{i}Y)\downarrow,\\
u_5&=&-\frac{1}{\sqrt{2}}(X+\mbox{i}Y)\downarrow\\
 u_6&=&Z\downarrow,
\end{eqnarray}

In the above basis, the $kp$ matrix reads
\begin{equation}
\frac{\hbar^2}{2m_o}\left[\begin{array}{cccccc}
 F & -K^*& -H^*&  0&   0& 0\\
 -K & F-2\Delta_2&   H&   0&   0&  \sqrt{2}\Delta_3\\
 -H &H^*&  L -\Delta_1-\Delta_2 &0 &  \sqrt{2}\Delta_3&  0\\
 0&0 &  0&   F&  -K&  H\\
  0&   0&   \sqrt{2}\Delta_3&  -K^* & F-2\Delta_2& -H^*\\
 0 & \sqrt{2}\Delta_3&   0&   H^*& -H&  L-\Delta_1-\Delta_2
\end{array}\right].
\end{equation}
Here,
\begin{eqnarray}
L &=& A_1k_z^2 + A_2(k_x^2 + k_y^2),\\
 T &=& A_3k_z^2 + A_4(k_x^2 +k_y^2),\\
  F &=& L + T,\\
k_+ &=& k_x + \mbox{i}k_y,\\
 K &=& A_5k_+^2,\\
 H &=& A_6k_+k_z,
\end{eqnarray}
where $\Delta_i$ and $A_i$ are the valence band splittings and $kp$
parameters, respectively. \ \\

p-d exchange matrix is given by
\begin{equation}
H_{pd} = 3B_G\left[\begin{array}{cccccc}
 b_xw_z & 0&      0&        0& b_xw_m&      0\\
 0 &   b_xw_z&   0 &      b_xw_m&    0 &       0\\
 0&     0  &  b_zw_z &     0&       0&        b_zw_m\\
 0&    b_xw_p&   0&      -b_xw_z&    0&         0\\
 b_xw_p & 0&      0 &        0&     -b_xw_z &   0\\
 0&      0&     b_zw_p&      0&     0&      -b_zw_z
\end{array}\right].
\end{equation}

\section{Numerical procedure}

We aim at evaluating the Helmholtz free energy $F_c$ at given hole
concentration $p$ as a function of the Mn magnetization $M$,
\begin{equation}
F_c(p,M) = -k_BT\int {\mbox{d}} \varepsilon N(\varepsilon)
\ln\left(1+\exp[-(\varepsilon(M) -\varepsilon_F(p,M))/k_BT]\right)
+ p\varepsilon_F(p,M),
\end{equation}
where the thermodynamic density-of-states $N(\varepsilon)=\partial
p/\partial\varepsilon$. If $T < 200$ K then in the studied cases
the hole liquid is degenerate,  so that $F_c$ assumes a simple
form,
\begin{equation}
F_c(p,M) = \int_0^p{\mbox{d}}p\prime \varepsilon(M,p\prime).
\end{equation}

Thus, in order to obtain $F_c(p,M)$ we have to integrate the
dependence of the partition function or of the Fermi energy on the
hole concentration for various $M$. The actual calculation proceeds
in a standard way. First, the length of the wave vectors
$k_i(\theta,\varphi)$ for each of six hole subbands is determined
by solving the inverse eigenvalue problem for given values of
$\varepsilon$,  polar angle $\theta$ and azimuth angle $\varphi$ in
the ${\bm k}$-vector space. Then, the hole concentration is
obtained by an integration of $\sum_i
k_i^3(\theta,\varphi)/24\pi^3$ over $\cos\theta$ and $\varphi$.

In order to calculate the contribution $M_c$ of the hole magnetic
moments to the total magnetization, the eigenvalue problem is
solved directly making it possible to determine the Gibbs
thermodynamic potential $G_c$,
\begin{equation}
G_c=-\sum_{i,n,k_z}k_BT(eB/2\pi\hbar)\ln\left(1+\exp[
-(\varepsilon_i(n,k_z,M)
-\varepsilon_F)/k_BT]\right).
\end{equation}

We calculate the absorption coefficient $\alpha$ for the two
circular polarizations $\sigma^{\pm}$, taking into account $k$
conserving electron transitions from the six valence band subbands,
index $i$, to the two spin branches of the conduction band, index
$j$, assuming that the hole liquid is strongly degenerate. In such
a model,\cite{Kane57,thanks}
\begin{equation}
\alpha^{\pm}= \frac{4\pi^2e^2P^2}{\hbar^2 c n_r\omega}
\sum_{i,j}\int_{-1}^1
\mbox{d}\cos\theta\int_0^{2\pi}\mbox{d}\varphi\frac{k_{ij}(\omega)m_r^{(ij)}(\omega)}
{8\pi^3\hbar^2}|M_{ij}^{\pm}|^2.
\end{equation}
The wave vectors $k_{ij}(\omega,\theta,\varphi)$ corresponding to
twelve possible transitions are determined by the energy
conservation and the position of the Fermi level. The joint density
of states effective mass corresponding to these $k_{ij}$ is given
by
\begin{equation}
m_r^{(ij)}=(1/m_e -
\frac{1}{\hbar^2k_{ij}}\frac{\partial\varepsilon_{i}}{\partial
k_{ij}})^{-1}.
\end{equation}
The matrix elements $M_{ij}^{\pm}(\theta,\varphi)$ for the two
light polarizations and involving electron transitions to the spin
down and spin up conduction subbands are given by
\begin{eqnarray}
M_{i1}^+&=& a_i^{(4)},\\
 M_{i2}^+&=& a_i^{(3)}/\sqrt{3} + \mbox{i}a_i^{(6)}\sqrt{2/3},\\
 M_{i1}^-&=& a_i^{(2)}/\sqrt{3} - \mbox{i}a_i^{(5)}\sqrt{2/3},\\
M_{i2}^-&=& a_i^{(1)},
\end{eqnarray}
where $a_i^{n}$ is the $n$-th component of the eigenvector
corresponding $i$-th valence subband at $k_{ij}(\theta,\varphi)$.

\twocolumn
\section{Material parameters}

In Tables I and II we summarize band structure parameters of parent
compounds employed for the evaluation of chemical trends in cubic
and wurzite magnetic semiconductors.
%
%

\begin{table}[p] \centering \caption[]{Material parameters of
selected cubic semiconductors and the values of the p-d exchange
energy $\beta N_o$. Except for the parameters for which references
are provided, the values of the lattice constant $a_o$, spin-orbit
splitting $\Delta_o$, and Luttinger parameters $\gamma_i$ are taken
from {\it Landolt-B\"orstein}.~\cite{Bimb82,Blac99,Bros82,Made87}}
\label{table:zinc}
\begin{tabular}{ccccccc}
 & $a_o$ ({\AA})&  $\Delta_o$ (eV)& $\gamma_1$& $\gamma_2$& $\gamma_3$&$\beta N_o$ (eV)\\
\hline Si  & 5.43& 0.044 & 4.285& 0.339&1.446&-1.35$^a$ \\ Ge  &
5.66 & 0.29  & 13.38&4.24&5.69&-1.20$^a$\\ AlP& 5.47 & 0.1$^a$ &
3.47 &0.06&1.15  &-1.33$^a$ \\ AlAs& 5.66& 0.275 & 3.25 &0.64&1.21
& -1.19$^a$ \\ GaN&4.50$^b$&0.018$^b$&2.463$^c$ &0.647$^c$
&0.975$^c$ &-2.37$^a$ \\ GaP& 5.45 & 0.080  & 4.05
&0.49&1.25&-1.34$^a$ \\ GaAs& 5.65& 0.34 & 6.85 & 2.1 &
2.9&-1.2$^d$ \\ GaSb & 6.09 & 0.76  & 13.3  &4.4&5.7&-0.96$^a$ \\
InP& 5.87& 0.108& 5.15& 0.94&1.62& -1.07$^a$ \\ InAs& 6.06& 0.38  &
20.4  &8.3&9.1&-0.98$^a$ \\ ZnS& 5.401& 0.070 & 1.77$^a$ & 0.30$^a$
&0.62$^a$ &-1.5$^e$ \\ ZnSe& 5.67 & 0.43& 2.95$^f$ &0.6$^f$
&1.11$^g$ &-1.3$^h$ \\ ZnTe&6.10& 0.91& 3.8& 0.72&1.3& -1.1$^i$ \\
CdTe&6.48& 0.95& 4.14& 1.09&1.62$^j$&-0.88$^j$\\
\end{tabular}
\end{table}
\ \\ \ \\ \ \\ \ \\ \ \\ \ \\ \ \\ \ \\ \ \\ \ \\ \ \\ \ \\ \ \\ \
\\ \ \\ \ \\ \noindent \small $^a$ the value determined from
Eq.~(\ref{eq:III-V})\\ \small $^b$ the value determined from the
data for the wurzite structure (Table \ref{table:wurzite})\\ \small
$^c$ K. Kim, W.L.R. Lambrecht, B. Segall, and M. van Schilfgaarde,
Phys. Rev. B {\bf 56}, 7363 (1997); in this paper the Dresselhaus
parameters $L, M, N$ are denoted as $A, B, C$, respectively; see
also, I. Stolpe, N. Puhlmann, H.-U. M\"uller, O. Portugall, M. von
Ortenberg, D. Schikora, D.J. As, B. Sch\"ottker, and R. Lischka,
Physica B {\bf 256-258}, 659 (1998)\\ \small $^d$ the value from
photoemission.\cite{Okab98}\\
 \small $^e$ the value determined from Eq.~(\ref{eq:II-VI})\\
\small $^f$ see the main body of the text \\ \small $^g$ H.W.
H\"olscher, A. N\"othe, and Ch. Uihlein, Phys. Rev. B {\bf 31},
2379 (1985)\\ \small $^h$magnetoreflectivity\cite{Twar84b} \\
\small $^i$magnetoreflectivity\cite{Twar84a} \\ \small $^j$
magnetoreflectivity\cite{Gaj79}
\\

\begin{table}[tbp]
\centering \caption[]{Material parameters of selected wurzite
semiconductors. The values of the lattice parameter $a_o = (\sqrt 3
a^2c)^{1/3}$ and the splittings $\Delta_i$ of the valence band at
the $\Gamma$ point in II-VI compounds are taken from {\it
Landolt-B\"orstein}.\cite{Blac99,Bros82} Theoretical studies
provide the parameters $\Delta_i$ and $A_i$ for GaN,\cite{Kim97}
InN,\cite{Yeo98}  and $A_i$ for ZnO, CdS, and CdSe.\cite{Vogl96}}
\label{table:wurzite}
\begin{tabular}{lccccc}
\small Material &GaN& InN& ZnO& CdS & CdSe\\
\hline
 $a_o$ ({\AA}) &4.503 &4.974 &4.567 &5.845 &6.078\\ $\Delta_1$ (eV) &
 0.0036 &0.017
&0.04 &0.030 &0.039\\ $\Delta_2$ (eV) &0.005  &0.001  &0.0  &0.022
&0.139\\ $\Delta_3$ (eV) &0.0059  &0.001  &0.0  &0.022 &0.139\\
$A_1$& -6.4 &-9.28 &-2.41  &-5.92 &-10.2 \\ $A_2$&-0.5  &-0.6
&-.044  &-0.70 &-0.76\\ $A_3$ & 5.9 &8.68  &2.11  &5.37 &9.53\\
$A_4$&-2.55&-4.34  &-1.06  &-1.82 &-3.2\\
$A_5$&-2.56 &-4.32 &-1.06
&-1.82 &-3.2\\
$A_6$&-3.06 &-6.08  &-1.51&-1.36&-2.31\\
 $\beta N_o$(eV) & -2.37$^a$ & -1.76$^a$ &-2.48$^b$ &-1.55$^c$ &-1.27$^d$\\
  \end{tabular}
\end{table}

\ \\ \ \\ \ \\ \ \\
\ \\ \ \\ \ \\ \ \\
\ \\ \ \\ \ \\ \ \\
 \noindent \small $^a$ Calculated from
Eq.~(\ref{eq:III-V})\\ $^b$ Calculated from Eq.~(\ref{eq:II-VI})\\
$^c$ interpretation of magnetooptical data\cite{Beno92}\\ $^d$
magnetoabsorption\cite{Arci86}\\


\newpage
\begin{thebibliography}{99}
\bibitem{Ohno92} H. Ohno, H. Munekata, T. Penney, S. von Moln\`ar,
 and L.L. Chang, Phys. Rev. Lett. {\bf 68}, 2664 (1992).
 \bibitem{Ohno96a} H. Ohno, A. Shen,  F. Matsukura, A.
Oiwa, A. Endo, S. Katsumoto, and Y. Iye, Appl. Phys. Lett.  {\bf
69} 363 (1996).
\bibitem{Haur97} A. Haury, A. Wasiela,  A. Arnoult, J. Cibert,
S. Tatarenko, T. Dietl, and  Y. Merle d'Aubign\'e, Phys. Rev. Lett.
{\bf 79}, 511 (1997).
\bibitem{Ferr99} D. Ferrand, J. Cibert, C. Bourgognon,
S. Tatarenko, A. Wasiela, G. Fishman, A. Bonanni, H. Sitter,
S. Kole\'snik, J. Jaroszy\'nski, A. Barcz, and T. Dietl, J. Crystal
Growth and Physica B, in press; e-print:
xxx.lanl.gov/cond-matt/9910131.
\bibitem{Kosh97} See, S. Koshihara, A. Oiwa, M. Hirasawa,
S. Katsumoto, Y. Iye, C. Urano, H. Takagi, and H. Munekata, Phys.
Rev. Lett. {\bf 78}, 4617 (1997).
 \bibitem{Ohno99b} Y. Ohno,  D.K. Young, B. Beschoten,
 F. Matskura, H. Ohno, and D.D. Awschalom, Nature {\bf 402}, 790 (1999).
\bibitem{Mats98} F. Matsukura, H. Ohno, A. Shen, and Y. Sugawara,
Phys. Rev. B {\bf 57}, R2037 (1998).
\bibitem{Diet00} T. Dietl, H. Ohno, F. Matsukura, J. Cibert,
and D. Ferrand, Science {\bf 287}, 1019 (2000).
\bibitem{Ohno98}H. Ohno, Science {\bf 281}, 951 (1998); J. Magn. Magn.
Mater. {\bf 200}, 110 (1999); J.K. Furdyna, P. Schiffer, Y. Sasaki,
S.J. Potashnik, and X.Y. Liu, in: {\sl Optical Properties of
Semiconductor Nanostructures}, edited by M.L. Sadowski, M.
Potemski, and M. Grynberg (Kluwer, Dordrecht, 2000) p. 211.
\bibitem{Cibe99} J. Cibert, P. Kossacki, A. Haury, D. Ferrand,
A. Wasiela, Y. Merle d'Aubign\'e, A. Arnoult, S. Tatarenko, and T.
Dietl, in Proceedings of 24th International Conference on Physics
of Semiconductors, Jerusalem, August 1998, edited by D. Gershoni
(World Scientific, Singapore), p.51.
\bibitem{Stor97} T. Story, Acta Phys. Polon. A {\bf 91}, 173 (1997).
 \bibitem{Shio98} R. Shioda, K. Ando, T. Hayashi, and M.
Tanaka, Phys. Rev. B {\bf 58}, 1100 (1998).
\bibitem{Balz85} A. Balzarotti, N. Motta, A. Kisiel,
M. Zimnal-Starnawska, M.T. Czy\.zyk, and M. Podg\'orny Phys. Rev. B
{\bf 31}, 7526 (1985).
\bibitem{Oiwa97} A. Oiwa, S. Katsumoto, A. Endo, M. Hirasawa, Y. Iye,
H. Ohno, F. Matsukura, A. Shen, Y. Sugawara, Solid State Commun.
{\bf 103}, 209 (1997).
\bibitem{Esch97} A. Van Esch, L. Van Bockstal, J. De Boeck, G.
Verbanck, A.S. van Steenbrgen, P.J. Wellmann, B. Grietens, R.
Bogaerts, F. Herlach, and G. Borghs, Phys. Rev. B {\bf 56}, 13103
(1997).
 \bibitem{Shim99} H. Shimizu, T. Hayashi, T. Nishinaga, M. Tanaka,
Appl. Phys. Lett. {\bf 74}, 398 (1999).
 \bibitem{Omiy00} T. Omiya, F. Matsukura, T. Dietl, Y. Ohno, T. Sakon,
M. Motokawa, and H. Ohno, Physica E, in press.
\bibitem{Walu88} W. Walukiewicz, Phys. Rev. B {\bf 37}, 4760 (1988).
\bibitem{Linn97} M. Linnarsson, E. Janzén, B. Monemar, M.
Kleverman, and A. Thildekvist, Phys. Rev. B {\bf 55}, 6938 (1997).
\bibitem{Aver87} N.S. Averkiev, A.A. Gutkin, E.B. Osipov, and M.A. Reshchikov, Fiz. Tekh.
Poluprovodn. {\bf 21}, 1847 (1987) [Sov. Phys. Semicond. {\bf 21},
 1119 (1987)].
\bibitem{Zhan88} F.C. Zhang and T.M. Rice Phys.
Rev. B {\bf 37}, 3759 (1988).
\bibitem{Beno92} C. Benoit \`a la Guillaume, D. Scalbert, and T.
Dietl, Phys. Rev. B {\bf 46}, 9853 (1992).
\bibitem{Bhat00} A.K. Bhattacharjee and C. Benoit \`a la Guillaume,
Solid State Commun. {\bf 113}, 17 (2000).
\bibitem{Okab98} J. Okabayashi, A. Kimura, O. Rader, T. Mizokawa,
A. Fujimori, T. Hayashi, and M. Tanaka, Phys. Rev. B {\bf 58},
R4211 (1998).
\bibitem{Okab99} J. Okabayashi, A. Kimura, T. Mizokawa, A. Fujimori,
T. Hayashi, and M. Tanaka, Phys. Rev. B {\bf 59}, R2486 (1999).
\bibitem{Diet94} for a review on DMS, see, e.g., T. Dietl, in
{\it Handbook on Semiconductors}, edited by T.S. Moss
(North-Holland, Amsterdam, 1994) vol.~3b, p. 1251.
\bibitem{Harr87} W.A. Harrison and G.K. Straub, Phys. Rev.
B {\bf 36}, 2695 (1987).
\bibitem{Akai98} H. Akai, Phys. Rev. Lett. {\bf 81}, 3002 (1998).
\bibitem{Paal91} M.A. Paalanen and R.N. Bhatt, Physica B {\bf 169},
153 (1991).
\bibitem{Sawi86} M. Sawicki, T. Dietl, J. Kossut, J. Igalson,
T. Wojtowicz, and W. Plesiewicz, Phys. Rev. Lett.  {\bf 56}, 508
(1986).
\bibitem{Glod94} P. G{\l}\'od, T. Dietl, M. Sawicki, and I. Miotkowski,
Physica B {\bf 194-196}, 995 (1994).
\bibitem{Diet97} T. Dietl, A. Haury, and Y. Merle d'Aubign\'e,
Phys. Rev. B {\bf 55}, R3347 (1997).
\bibitem{Diet99} T. Dietl, J. Cibert, D. Ferrand, and
Y.~Merle d'Aubign\'e, Mater. Sci. Engin. B {\bf 63}, 103 (1999),
and references therein.
\bibitem{Jung99} T. Jungwirth, W.A. Atkinson, B.H. Lee, and
A.H. MacDonald, Phys. Rev. B {\bf 59}, 9818 (1999).
  \bibitem{Beli94}  D. Belitz and T.R. Kirkpatrick, Rev. Mod. Phys.
{\bf 57}, 287 (1994), and references therein.
\bibitem{Szcz99b} J. Szczytko, K. \'Swi{\c{a}}tek, M. Palczewska,
A. Twardowski, T. Hayashi, M. Tanaka, and K. Ando, Phys. Rev. B
{\bf 60}, 8304 (1999) .
\bibitem{Szcz99a} J. Szczytko, W. Mac, A. Twardowski, F. Matsukura,
and H. Ohno, Phys. Rev. B {\bf 59}, 12935 (1999).
\bibitem{Besc99} B. Beschoten, P.A. Crowell, I. Malajovich,
 D.D. Awschalom, F. Matskura, A. Shen, and H. Ohno,
 Phys. Rev. Lett. {\bf 83}, 3073 (1999).
 \bibitem{Bir74} G.L. Bir and
G.E. Pikus, {\it Symmetry and Strain-Induced Effects in
Semiconductors} (Wiley, New York, 1974).
\bibitem{Gaj78} J.A. Gaj, J. Ginter, and R.R. Ga{\l}{\c{a}}zka,
Phys. Status Solidi (b)  {\bf 89}, 655 (1978).
 \bibitem{Nish97} Y. Nishikawa, Y.
Satoh, and J. Yoshino, in: {\it Extended Abstracts of the 2nd
Symposium of Spin-Related Phenomena in Semiconductors}, Sendai,
January, 1997, p. 122, unpublished.
\bibitem{Shap86} Y. Shapira, S. Foner, P. Becla, D.N. Domingues,
M.J. Naughton, and J.S. Brooks, Phys. Rev. B {\bf 33}, 356 (1986).
\bibitem{Twar84a} A.Twardowski, P. \'Swiderski, M. von Ortenberg, and R.
Pauthenet, Solid State Commun. {\bf 50}, 509 (1984).
\bibitem{Ferr00} D. Ferrand, J. Cibert, C. Bourgognon,
S. Tatarenko, A. Wasiela, G. Fishman, T. Andrearczyk,
S. Kole\'snik, J. Jaroszy\'nski,  T. Dietl, B. Barbara, and D.
Dufeu, unpublished; J. Appl. Phys. {\bf 87}, 6451 (2000).
\bibitem{Wolf96} P.A. Wolff, R.N. Bhatt, and A.C.
Durst, J. Appl. Phys. {\bf 79}, 5196 (1996).
\bibitem{Bhat99}  R.N. Bhatt and Xin Wan, Intern. J. Modern Phys.
C {\bf 10}, 1459 (1999).
\bibitem{Zene50} C. Zener, Phys. Rev. {\bf 81}, 440 (1950); ibid. {\bf
83} 299 (1950); a similar model for the nuclear ferromagnetism was
developed by H. Fr\"ohlich and F.R.N. Nabarro, Proc. Roy. Soc.
(London) A {\bf 175}, 382 (1940); see also,  e.g., P. Leroux-Hugon,
in: {\it New Developments in Semiconductors}, edited by P.R.
Wallace, R. Harris, and M.J. Zuckermann (Noordhoff, Leyden, 1973)
p.~63; M.A. Krivoglaz, Usp. Fiz. Nauk  {\bf 111} (1973) 617 [Sov.
Phys. Usp.  {\bf 16}, 856(1974)]; E.A. Pashitskii and S.M.
Ryabchenko, Fiz. Tver. Tela {\bf 21}, 545 (1979) [Sov. Phys. Solid
State {\bf 21}, 322 (1979)].
\bibitem{Blin97} J. Blinowski, P. Kacman, and J.A. Majewski,
in: {\it High Magnetic Fields in Semiconductor Physics II}, eds. G.
Landwehr, W. Ossau (World Scientific, Singapore 1997), p. 861.
 \bibitem{Abol00} R. Abolfath, J. Brum, and A.H. MacDonald,
 unpublished.
\bibitem{Szym78} W. Szyma\'nska and T. Dietl,
 J. Phys. Chem. Solids {\bf 39}, 1025 (1978).
 \bibitem{Kitt68} C. Kittel, {\it Solid State Physics}, vol. 22,
 edited by F. Seitz, D. Turnbull, and H. Ehrenreich (Academic Press,
 New York, 1968), p.~1.
\bibitem{Lars81} U. Larsen,
Phys. Lett. {\bf 85A}, 471 (1981).
\bibitem{Koss00} P. Kossacki, D.
Ferrand, A. Arnoult, J. Cibert, S. Tatarenko, A. Wasiela, Y. Merle
d'Aubign\'e, K. \'Swi{\c{a}}tek, M. Sawicki, J. Wr\'obel, W.
Bardyszewski, and T. Dietl, Physica E {\bf 6}, 709 (2000).
 \bibitem{Fish72} M.E. Fisher, S.-k. Ma, and B.G. Nickel,
Phys. Rev. Lett.  {\bf 29}, 917 (1972).
\bibitem{Yeom93} See, J.M. Yeomans, {\it Statistical
Mechanics of Phase Transitions}, (Oxford University Press, Oxford,
1993), p. 57.
 \bibitem{Nish98} K. Nishizawa and O. Sakai, in: {\it Extended Abstacts, 4th
Symposium on the Spin-Related Phenomena in Semiconductors}, Sendai,
December, 1998, p. 140, unpublished.
\bibitem{Boss00} M.A. Boselli, A. Ghazali, and I.C. da Cunha
Lima, e-print: xxx.lanl.gov/cond-matt/0002254.
\bibitem{Koni00} J. K\"onig, H.-H. Lin, and
A.H. MacDonald, Phys. Rev. Lett. {\bf 84}, 5628 (2000).
\bibitem{Shir98} M. Shirai, T. Ogawa,  I. Kitagawa, and N. Suzuki
J. Magn. Magn, Mater. {\bf 177-181}, 1383 (1998).
\bibitem{Kato99} R. Kato and H. Katayama-Yoshida,
 in: {\it Extended Abstracts of the 5th Symposium of Spin-Related
Phenomena in Semiconductors}, Sendai, December, 1999, p. 233,
unpublished.
 \bibitem{Egge95} P.T.J. Eggenkamp, H.J.M. Swagten, T. Story,
 V.I. Litvinov, C.H.W. Sw\"uste, and W.J.M. de Jonge, Phys.
 Rev. B {\bf 51}, 15~250 (1995).
\bibitem{Ohno99a} H. Ohno, F. Matsukura, T. Omiya, and N. Akiba,
J. Appl. Phys. {\bf 85}, 4277 (1999).
\bibitem{Ohno96b} H. Ohno, F. Matsukura, A. Shen,  Y. Sugawara,
A. Oiwa, A. Endo, S. Katsumoto, and Y. Iye, in: {\it Proc. 23rd
International Conference on the Physics of Semiconductors}, Berlin
1996, edited by M. Scheffler and R. Zimmermann (World Scientific,
Singapore, 1996) p. 405.
\bibitem{Hube98} A. Hubert and R. Schafer, {\it Magnetic Domains}
(Springer, Berlin, 1998).
\bibitem{Bimb82} D. Bimberg et al., {\it  Landolt-B\"orstein}, New-Series,
vol.~III/17a, edited by O. Madelung, M. Schulz, and W. Weiss
(Springer, Berlin, 1982).
\bibitem{Furd88b} see, {\it Diluted Magnetic Semiconductors},
edited by J.K. Furdyna and J. Kossut,  Semiconductors and
Semimetals, vol. 25 (Academic Press, Boston, 1988).
\bibitem{Szcz96} J. Szczytko, W. Mac, A. Stachow, A. Twardowski,
P. Becla, and J. Tworzyd{\l}o, Solid State Commun. {\bf 99}, 927
(1996).
\bibitem{Ando98} K. Ando, T. Hayashi, M. Tanaka, and A.
Twardowski, J. Appl. Phys. {\bf 83}, 6548 (1998).
\bibitem{Cibe98} J. Cibert, P. Kossacki, A. Haury, A. Wasiela,
Y. Merle d'Aubign\'e, T. Dietl, A. Arnoult, and S. Tatarenko, {\it
Proc. Intl. Conf. II-VI Compounds}, Grenoble 1997, J. Crystal
Growth {\bf 184/185}, 898 (1998).
\bibitem{Case76}H.C. Casey Jr. and F.F. Stern, J. Appl. Phys.
 {\bf 47}, (1976) 631.
 \bibitem{Case75} H.C. Casey Jr., D.D. Sell, and K.W. Wecht,
 J. Appl. Phys. {\bf 46}, 250 (1975).
\bibitem{Gaj79} J.A. Gaj, R. Planel, and G. Fishman, Solid State
Commun. {\bf 29}, 435 (1979).
\bibitem{Lank96} S. Lankes, M. Meier, T. Reisinger, and W. Gebhard,
J. Appl. Phys. {\bf 80}, 4049 (1996).
 \bibitem{Hols85} H.W.
H\"olscher, A. N\"othe, and Ch. Uihlein, Phys. Rev. B {\bf 31},
2379 (1985).
\bibitem{Vogl96} The parameters $\gamma_i$ zinc-blende
structure ZnS and ZnSe and $A_i$ wurzite ZnO, CdS, and CdSe were
obtained by fitting the corresponding $kp$ model to the results of
band structure computations given in D. Vogel, P. Kr\"uger, and J.
Pollmann, Phys. Rev. B {\bf 54}, 5495 (1996).
\bibitem{Veng79} H. Venghaus, Phys. Rev. B {\bf 19}, 3071 (1979).
 \bibitem{Mizo99} T.
Mizokawa, J. Okabayashi, T. Nambu, and A. Fujimori, in: {\it
Extended Abstracts of the 5th Symposium of Spin-Related Phenomena
in Semiconductors}, Sendai, December 1999, unpublished.
\bibitem{Kats98} S. Katsumoto, A. Oiwa, Y. Iye, H. Ohno, F.
Matsukura, A. Shen, and Y. Sugawara, phys. status solidi (b) {\bf
205}, 115 (1998).
 \bibitem{Blac99} R. Blachnik et
al., {\it Landolt-B\"orstein}, New Series, vol. 41B, edited by U.
R\"ossler (Springer-Verlag, Berlin, 1999).
\bibitem{Twar84b} A. Twardowski, M. von
Ortenberg, M. Demianiuk, and R. Pauthenet, Solid State Commun. {\bf
51}, 849 (1984).
\bibitem{Wojt86} T. Wojtowicz, T. Dietl, M. Sawicki, W. Plesiewicz,  and J.
Jaroszy\'nski, Phys. Rev. Lett. {\bf 56}, 2419 (1986).
 \bibitem{Chmi87} M. Chmielowski, T. Dietl, P.
Sobkowicz,  and F. Koch, in: Proc. 18th Int. Conf. Physics of
Semiconductors, Stockholm 1986, ed. O. Engstr\"om (World
Scientific, Singapore, 1987) p. 1787.
 \bibitem{Hira99} H. Hirakawa,
S. Katsumoto, Y. Hashimoto, and Y. Iye, in: {\it Extended Abstracts
of the 5th Symposium of Spin-Related Phenomena in Semiconductors,
Sendai}, December, 1999, p. 174, unpublished.
 \bibitem{Naga99} Y. Nagai, K. Nagasaka, H. Nojiri, M. Motokawa,
 F. Matsukura, and H. Ohno,
 in: {\it Extended Abstracts of the 5th Symposium of Spin-Related
Phenomena in Semiconductors}, Sendai, December, 1999, p. 176,
unpublished.
 \bibitem{Kama99} T. Kamatani and H. Akai, in: {\it Extended Abstacts, 5th
Symposium on the Spin-Related Phenomena in Semiconductors}, Sendai,
December, 1999, p. 18, unpublished.
\bibitem{Good61} R.R. Goodman, Phys. Rev. {\bf 122}, 397 (1961).
 \bibitem{Kane57} E.O. Kane, J. Phys.
Chem. Solids {\bf 1}, 249 (1957).
\bibitem{thanks} We thank James J. Niggemann for his assistance
in the derivation of momentum matrix elements.
 \bibitem{Bros82} I. Broser et al., {\it Landolt-B\"orstein}, New
Series, vol. 17b, edited by O. Madelung, M. Schulz, and W. Weiss,
 (Springer-Verlag, Berlin, 1982).
 \bibitem{Made87} O. Madelung, W.
von der Osten, U. R\"ossler, {\it Landolt-B\"orstein}, New Series,
vol. 22a, edited by O. Madelung (Springer-Verlag, Berlin, 1987).
\bibitem{Kim97} K. Kim,
W.L.R. Lambrecht, B. Segall, and M. van Schilfgaarde, Phys. Rev. B
{\bf 56}, 7363 (1997); similar values of the $kp$ parameters $A_i$
  M. Suzuki, T. Uenoyama, and A. Yanase,
Phys. Rev. B {\bf 53}, 8132 (1995); J.A. Majewski, M. St\"adele,
and P. Vogl, in: III-V Nitrides, edited by T. Moustakas, B.
Monemar, I. Akasaki, and F. Ponce, {\it Materials Research Society
Symposium Proceedings No 449}, (MRS, Pittsburg, 1997); for the
review see, {\it Properties, Processing and Application of GaN and
Related Semiconductors}, edited by J.H. Edgar, S. Strite, I.
Akasaki, H. Amano, and C. Wetzel, An INSPEC Publication, pp.
153-207.
\bibitem{Yeo98} Y.C. Yeo, T.C.
Chong, and M.F. Li, J. Appl. Phys.
 {\bf 83}, 1429 (1998).
\bibitem{Arci86} M. Arciszewska and M.
Nawrocki, J. Phys. Chem. Solids {\bf 47}, 309 (1986).

\end{thebibliography}
\end{document}